\documentclass[traditabstract,longauth]{aa}  
\usepackage{graphicx}
\usepackage{txfonts}
\usepackage{lipsum}
\usepackage{subcaption}         
\usepackage{lscape}             
\usepackage{placeins}           
                                
\newcommand{\sersic}{S\'{e}rsic}
\newcommand{\hi}{{\rm H}{\textsc i}}
\newcommand{\nii}{[\rm N\textsc{ii}]}
\newcommand{\hii}{{\rm H}{\textsc{ii}}}
\newcommand{\oiii}{[\rm O\textsc{iii}]}

\usepackage[colorlinks=true, linkcolor=blue, citecolor=blue, urlcolor=blue]{hyperref}
\usepackage{float}
\let\linenumbers\relax
\let\endlinenumbers\relax

\begin{document}

   \title{AVID: A Near-Major Post-Merger of Late-Type Dwarfs beneath a Regularly Rotating $\hi$ Disk (VCC 693)}
   
    \author{Fujia Li\inst{1,2,3}
        \and Hong-Xin Zhang\inst{1,2}\fnmsep\thanks{hzhang18@ustc.edu.cn}
        \and Elias Brinks\inst{4}
        \and Se-Heon Oh\inst{5,6}\fnmsep\thanks{seheon.oh@sejong.ac.kr}
        \and Rory Smith\inst{7,8}
        \and Zesen Lin\inst{9}
        \and Weibin Sun\inst{1,2}
        \and Yu-Zhu Sun\inst{1,2}
        \and Tie Li\inst{1,2}
        \and Minsu Kim\inst{5,6}
        \and Jaebeom Kim\inst{5,6}
        \and Lijun Chen\inst{1,2}
        \and Lanyue Zhang\inst{1,2}
        \and Patrick C\^ot\'e\inst{10}
        \and Alessandro Boselli\inst{11}
        \and Pierre-Alain Duc\inst{12}
        \and Laura Ferrarese\inst{10}
        \and Matteo Fossati\inst{13}
        \and Stephen Gwyn\inst{10}
        \and Xu Kong\inst{1,2,3}\fnmsep\thanks{xkong@ustc.edu.cn}
        \and Sanjaya Paudel\inst{14,15}
        \and Eric W. Peng\inst{16}
        \and Thomas H. Puzia\inst{17}
        \and Rubén Sánchez-Janssen\inst{18}
        \and Matthew Taylor\inst{19}
        \and Yinghe Zhao\inst{20,21}
        }
        
    \institute{Department of Astronomy, University of Science and Technology of China, Hefei 230026, China 
    \and School of Astronomy and Space Science, University of Science and Technology of China, Hefei 230026, China 
    \and Institute of Deep Space Sciences, Deep Space Exploration Laboratory, Hefei 230026, China 
    \and Centre for Astrophysics Research, University of Hertfordshire, College Lane, Hatfield, AL10 9AB, UK 
    \and Department of Astronomy and Space Science, Sejong University, Seoul 05006, Republic of Korea 
    \and Department of Physics and Astronomy, Sejong University, Seoul 05006, Republic of Korea 
    \and Departamento de Fisica, Universidad Tecnica Federico Santa Maria, Avenida España 1680, Valparaíso, Chile 
    \and Millenium Nucleus for Galaxies (MINGAL), Valparaíso, Chile 
    \and Department of Physics, The Chinese University of Hong Kong, Shatin, N.T., Hong Kong S.A.R., China 
    \and National Research Council of Canada, Herzberg Astronomy and Astrophysics, 5071 West Saanich Road, Victoria, BC V9E 2E7, Canada 
    \and Aix-Marseille Univ., CNRS, CNES, LAM, Marseille, France 
    \and Universit\`e de Strasbourg, CNRS, Observatoire astronomique de Strasbourg (ObAS), UMR 7550, F-67000 Strasbourg, France 
    \and Universit\`a di Milano-Bicocca, Piazza della Scienza 3, 20100 Milano, Italy 
    \and Department of Astronomy, Yonsei University, Seoul, 03722, Republic of Korea 
    \and Center for Galaxy Evolution Research, Yonsei University, Seoul, 03722, Republic of Korea 
    \and NSF NOIRLab, 950 N. Cherry Avenue, Tucson, AZ 85719, USA 
    \and Instituto de Astrofísica, Pontificia Universidad Católica de Chile, 7820436 Macul, Santiago, Chile 
    \and UK Astronomy Technology Centre, Royal Observatory, Blackford Hill, Edinburgh, EH9 3HJ, UK 
    \and University of Calgary, 2500 University Drive NW, Calgary, Alberta, T2N 1N4, Canada 
    \and Yunnan Observatories, Chinese Academy of Sciences, Kunming 650011, People's Republic of China 
    \and State Key Laboratory of Radio Astronomy and Technology, National Astronomical Observatories, Chinese Academy of Sciences, Beijing 100101, China 
}
    \date{February, 2026}

  \abstract
{
  On the periphery of galaxy clusters, the combination of moderately high galaxy number densities and modest velocity dispersions favours interactions and mergers that influence galaxy evolution prior to cluster infall.
  Observational studies of this important phase of galaxy evolution---particularly in the dwarf galaxy regime---are still rare.
  This study presents a high-resolution case study of VCC 693 (stellar mass $\rm \sim 2.8\times10^{8}\,M_{\odot}$), a near-major merger remnant in the outskirts of the Virgo cluster, using observations from the Atomic gas in Virgo Interacting Dwarf galaxies (AVID) project.
  We explore the consequences of the merger on the star formation and structural evolution of VCC 693, based on a joint analysis of high-resolution VLA and high-sensitivity FAST $\hi$ emission line observations, optical broadband images, narrow-band H$\alpha$ images, and optical spectra.
  We also employ hydrodynamical simulations of dwarf-dwarf mergers to aid in interpreting the observations of VCC 693.
  Our analysis favors a near-major merger between two late-type dwarf galaxies with a stellar mass ratio of approximately 3:1-4:1, in which one of the progenitors is likely relatively gas-poor (i.e., a damp merger).
  The optical appearance of VCC 693 is dominated by complex tidal structures spanning the whole system, whereas the $\hi$ gas has settled to a regular rotating disk with normal surface density profile. 
  Compared to dwarfs of similar mass, the star formation and gas-phase metallicity are moderately enhanced in the centre. 
  The global star formation rate, $\hi$ gas content, and $\hi$-to-optical size ratio of VCC 693 are broadly consistent with those of typical late-type dwarfs of similar mass, although they fall at the lower side of the distributions.  
  By decomposing the $\hi$ rotation curve into baryonic and dark matter component, we find that the dark matter halo is characterized by an unusually high concentration or core density.
  This implies that the dark matter halo may have relaxed into a more centrally peaked distribution following the merger event. 
  Together with two other recent studies of AVID post-merger systems, our findings reinforce the emerging view that even major mergers between dwarfs can produce remnants whose overall stellar structures---apart from tidal features---are indistinguishable from those of ordinary dwarfs. 
  Our results also suggest that the diverse environmental effects experienced by galaxies in cluster outskirts can promote damp or mixed mergers, which constitute an integral part of galactic pre-processing.
}


   \maketitle

\section{Introduction}
In the standard $\Lambda$CDM model, dwarf galaxies are the most abundant type of galaxy in the Universe, growing through star formation and merger processes. 
Of particular importance in the study of dwarf galaxies is their neutral hydrogen ($\hi$) gas, which serves as the fuel for star formation and plays a key role in driving galaxy evolution. 
Dwarf galaxies are also considered local analogues of primordial galaxies in the early Universe due to their low metal content \citep[e.g.,][]{Tolstoy2009,Zhao2013}. 
However, observations of $\hi$ gas in dwarf galaxies have been limited by their small size and the required sensitivity. 
Over the past two decades, several high-resolution $\hi$ surveys of dwarf galaxies in the local Universe (within $\sim$10 Mpc) have been conducted to derive their spatially resolved $\hi$ properties and to examine the morphology and dynamics of star-forming dwarf galaxies. 
These include THINGS \citep[The $\hi$ Nearby Galaxy Survey;][]{Walter2008}, FIGGS \citep[The Faint Irregular Galaxies GMRT Survey;][]{Begum2008}, SHIELD \citep[The Survey of $\hi$ in Extremely Low-mass Dwarfs;][]{Cannon2011}, and LITTLE THINGS \citep[Local Irregulars That Trace Luminosity Extremes in The $\hi$ Nearby Galaxy Survey;][]{Hunter2012}.
These surveys, along with some observations of individual targets, have studied their baryonic matter and dark matter distributions through their kinematics and explored how the star formation cycle is associated with the interstellar medium (ISM) content in dwarf galaxies \citep[e.g.,][]{Oh2015,Cannon2016,Hunter2021}.

In addition to the star formation cycle within the galaxy itself, environmental effects also play an important role in driving the evolution of galaxies, particularly through processes such as ram pressure stripping \citep[RPS, e.g.,][]{Gunn1972,Dickey1991,Quilis2000,Boselli2022}, tidal interactions \citep[e.g.,][]{Brosch2004,WangJ2023}, harassment \citep[e.g.,][]{Moore1996}, strangulation \citep[e.g.,][]{Larson1980,Bekki2002}, and galaxy mergers \citep[e.g.,][]{Toomre1972,Mihos1996,Cibinel2019}.
Dwarf galaxies, at the bottom of the hierarchical formation of galaxy groups and clusters, have weakly bound ISM due to their low gravitational potential, making them excellent targets for investigating any environmental effects \citep{Boselli2008,Grossi2015}.
Most studies of galaxy mergers have focused on massive galaxies, and it is only in the past decade that mergers between dwarfs have received more attention in both simulations \citep[e.g.,][]{Deason2014,Pearson2016} and observations \citep[e.g.,][]{Stierwalt2015,Paudel2017,Paudel2018,Kado-Fong2020,Zhang2020,Micic2023,Micic2024,Chauhan2025}.
However, we still lack a comprehensive understanding of the diversity of dwarf mergers and how they affect the properties of the ISM and star formation activities, especially near or within dense galaxy cluster environments.
In the outskirts of galaxy clusters, the moderately high galaxy number density and modest velocity dispersion favor interactions and mergers, which serve as an important pre-processing mechanism for infalling galaxies \citep{Zabludoff1998,Fujita2004}.

The AVID (Atomic Gas in Virgo Interacting Dwarfs; Zhang et al., in prep) survey was designed to address the above questions.
It was granted $\sim$ 176 hours with the National Radio Astronomy Observatory (NRAO\footnote{The National Radio Astronomy Observatory is a facility of the National Science Foundation operated under cooperative agreement by Associated Universities, Inc.}) Karl G. Jansky Very Large Array \citep[VLA,][]{Perley2011} in B, C, and D configurations to obtain high-resolution $\hi$ observations of 14 post-merger or highly disturbed dwarfs (with another 21 normal dwarfs within the fields of view imaged) in the Virgo cluster.
The AVID sample comprises systems with single-dish $\hi$ detections, selected from a relatively complete set of dwarf merger candidates identified through optical imaging.
Additionally, deep on-the-fly $\hi$ mapping observations were carried out for the AVID galaxies with the Five-Hundred-meter Aperture Spherical radio Telescope \citep[FAST;][]{Nan2011} as part of the AVID project, providing complementary information on low column density $\hi$ gas that may be missed by the VLA observations.
Together with other auxiliary multi-wavelength data, AVID represents the first project dedicated to studying tidal interactions and mergers of late-type dwarf galaxies in the outskirts of a galaxy cluster.

Based on the AVID project, our previous works presented the first case study of late-type major mergers that experienced substantial environmental stripping \citep[VCC 479;][]{SunW2025} and a thorough exploration of the nature and fate of a mysterious almost-dark cloud \citep[AGC 226178;][]{SunY2025}. 
In this paper, we present our comprehensive analysis of VCC 693, which is a unique post-merger system in the AVID sample and exhibits the most complex tidal features---including streams, tails, rings, and off-centre bars, as revealed by the deep optical imaging data from the Next Generation Virgo Cluster Survey \citep[NGVS,][]{Ferrarese2012}.
It is located on the outskirts of the Virgo cluster, with a velocity offset of 1069 $\rm km\,s^{-1}$ and a projected distance of 2.2 Mpc (7.4$^{\circ}$) from the cluster centre, M87.
The $\hi$ properties of VCC 693 have only been studied using single-dish observations, giving an $\hi$ mass of $\log M_{\hi} = 8.27\pm 0.01\,\rm M_{\odot}$ and an $\hi$ deficiency parameter of $\rm Def_{\hi}=0.27$ from \citet{Grossi2015}, suggesting only mild environmental effects.
However, pressure-based analyses, which compared the current pressure experienced from the intracluster medium with the pressure required to strip its $\hi$ gas, classified VCC 693 as a stripped galaxy \citep{Koppen2018,Minchin2022}.
By combining high-resolution $\hi$ data from interferometric observations with multi-wavelength data, we aim to understand how environmental effects and/or galaxy merger events have shaped the mass distribution, kinematic structure, star formation activity, and other relevant properties of VCC 693.
Table \ref{tab:property} presents some basic properties of VCC 693, either from the literature or derived in this work.

This paper is organized as follows. 
We describe our $\hi$ observations and other multi-wavelength data in Section \ref{sec: data}. 
We present the stellar morphology and environment of VCC 693 in Section \ref{sec: stellar properties}.
In Section \ref{sec:stellar and metallicity}, we use several analytical techniques on optical images to better display the stellar distribution, and subsequently, we analyze the star formation and metallicity properties of VCC 693.
In Section \ref{sec: HI properties} and Section \ref{sec: HI kinematics}, we discuss the $\hi$ gas distribution and kinematics of VCC 693, respectively.
We combine simulation and observation to discuss the formation and evolution history of VCC 693 in Section \ref{sec: Discussion}.
Finally, our main conclusions are presented in Section \ref{sec: Summary}.
Throughout this work, we adopt a distance of 16.5 Mpc (the average distance of the Virgo cluster) for VCC 693 \citep{Gavazzi1999,Mei2007,Auld2013,Cantiello2024}. 
Based on the baryonic Tully-Fisher relation from \citet{Lelli2019}, we also estimate a fiducial distance of $\sim21$ Mpc from its rotation velocity, which is consistent with 16.5 Mpc within the uncertainties, supporting our choice.

\begin{table}
\caption{Properties of VCC 693.\label{tab:property}}
\centering
\tabcolsep=3pt
\renewcommand\arraystretch{1.5}
\begin{tabular}{lcc}
\hline\hline
Property & Value & Reference \\
\hline
Optical Right Ascension (J2000) & 12$^h$24$^m$03.2$^s$ & 1\\
Optical Declination (J2000) & +05$^\circ$10$^\prime$50$^{\prime\prime}$ & 1\\
$\hi$ Right Ascension (J2000) & 12$^h$24$^m$03.3$\pm$0.1$^s$ & This work\\
$\hi$ Declination (J2000) & +05$^\circ$10$^\prime$47$\pm$2$^{\prime\prime}$ & This work\\
Adopted distance (Mpc) & 16.5 & 2, 3\\
Systemic velocity ($\rm km\,s^{-1}$) & 2,050$\pm1$& This work\\
Stellar mass ($\rm M_{\odot}$) & $2.8\pm0.2 \times 10^{8}$ & 4\\
$\hi$ mass ($\rm M_{\odot}$) & $1.6\pm0.2 \times 10^{8}$ & This work \\
Dust mass ($\rm M_{\odot}$) & $3.5\pm0.5 \times 10^{5}$ & 5\\
($g-i$) color (mag) & 0.43$\pm0.04$ & This work \\
Gas-phase metallicity & 8.52$\pm0.16$ & This work\\
SFR$\rm _{H\alpha}$ ($\rm M_{\odot}$ yr$^{-1}$) & 0.0196 & 4 \\
Total IR luminosity ($\rm L_{\odot}$) & $2.1\pm0.1 \times 10^{8}$ & 6 \\
Optical position angle ($^\circ$) & 327$\pm 4$ & This work\\
Optical inclination ($^\circ$) & 41$\pm 1$ & This work\\
$\hi$ position angle ($^\circ$) & 357$\pm3$ & This work\\
$\hi$ inclination ($^\circ$) & 45$\pm1$ & This work\\
Bar length from $\varepsilon$ profile (kpc) & 2.62 & This work\\
\hline
\end{tabular}
\tablefoot{Reference: (1)\citet{Abazajian2009}; (2) \citet{Auld2013}; (3) \citet{Gavazzi1999}; (4) \citet{Boselli2023} (5) \citet{Grossi2015}; (6) \citet{Minchin2022}}
\end{table}

\section{Observations and data reduction}\label{sec: data}
\subsection{VLA observations and calibration}
As part of the AVID project, VCC 693 was observed in the L band with VLA in 2021 (proposal ID: 21A-231, PI: Hong-Xin Zhang).
The total on-source time of VCC 693 is 4.6 hours for B-array (Dec 2021), 1.93 hours for C-array (June 2021), and 1.03 hours for D-array (April 2021) configurations.
A 16 MB bandwidth and a 1.67 $\rm km\,s^{-1}$ channel width are used for the $\hi$ line observations. 
$\hi$ emission line calibration and imaging were performed using the \texttt{CASA} software package \citep{McMullin2007}.
Particularly, by applying a cleaning mask in \texttt{TCLEAN}, the combined BCD data sets were de-convolved down to 1 $\sigma$ level. 
More details about the data reduction can be found in Zhang et al. (2025, in prep).
We created two versions of $\hi$ cubes by using different weighting schemes of the {\em uv} datasets in \texttt{TCLEAN}: one using ``natural weighting'', which assigns equal weight to all {\em uv} data points, and the other using ``Briggs weighting'' with the robust parameter set to $r=0.5$ (hereafter referred to as robust weighting).
The data cubes have a spectral resolution of 3.4 $\rm km\,s^{-1}$ after applying Hanning smoothing.
For the natural-weighted data cube, the resulting synthesized beam size is $13.^{\prime\prime}4\times9.^{\prime\prime}9$ and the rms noise is 0.52 mJy beam$^{-1}$. 
The robust-weighted data cube has a beam size of $7.^{\prime\prime}9\times6.^{\prime\prime}4$ and an rms noise of 0.59 mJy beam$^{-1}$.
These correspond to a 3$\sigma$ column density limit of $\rm 7.5\times10^{19}\,cm^{-2}$ and $\rm 2.2\times10^{20}\,cm^{-2}$ for the natural-weighted and robust-weighted data cubes, respectively, assuming a line width of 10 $\rm km\,s^{-1}$.

For the final data cubes, we created masks and a suite of two-dimensional moment maps, i.e. $\hi$ integrated intensity maps ({\sc moment0}), intensity-weighted mean $\hi$ velocity fields ({\sc moment1}), and velocity dispersion maps ({\sc moment2}), by using the Source Finding Application \citep[\texttt{SoFiA-2},][]{Serra2015,Westmeier2021}\footnote{\url{https://gitlab.com/SoFiA-Admin/SoFiA-2/-/tree/master?ref_type=heads}}.
We apply the \texttt{smooth and clip} algorithm with a threshold of 4$\sigma$, and smoothing kernels of 0 and 1.5 times the beam size for spatial kernels and 0, 3, 11, and 23 channels in the velocity direction.
These parameters are empirically determined.
Furthermore, the {\sc moment1} and {\sc moment2} maps are generated using only pixels with a signal-to-noise ratio (S/N) above 3.

\subsection{FAST}
Deep FAST $\hi$ observations of VCC 693 were carried out on 2022 August 12, 20, 30, and September 13 as part of the AVID project (proposal ID: PT2022\_0005, PI: Hong-Xin Zhang).
FAST is the largest single-dish radio telescope in the world, offering higher sensitivity than ALFALFA with its 500 m aperture (300 m illuminated at any instant) \citep{Nan2011,Jiang2019,Qian2020}.
Its L-band 19-beam receivers have a frequency range of 1.05$\sim$1.45 GHz with four polarization modes, and the FWHM of each raw beam is $\sim$2$^\prime$.9 at a frequency of 1.42 GHz.
The observations were conducted using the Multibeam on-the-fly mode, with a 23$^\circ$.4/53$^\circ$.4 (horizontal/vertical) rotation, 21$^\prime$.66 scanning separation, a 15$\rm ^{\prime \prime}\,s^{-1}$ scanning speed, covering approximately a uniform 30 by 30 arcmin sky area centred around VCC 693.
To calibrate the antenna temperature, the observation injected the signal of a 10 K noise diode for 1 second every 60 seconds.
The data were recorded by the Spec(W+N) backend with a sampling time of 0.5 s over 500 MHz and 65,536 channels, at a frequency and velocity resolution of 7.6 kHz (1.6 km s$^{-1}$) at 1.4 GHz.

The total on-source mapping time for a single pair of vertical and horizontal scanning observation of VCC 693 is about 16.1 minutes (64.4 minutes for all 4 observations).
The FAST data reduction and calibration were carried out with the $\hi$FAST pipeline\footnote{\url{https://hifast.readthedocs.io/en/v1.3/}} developed by \citet{Jing2024}.
After performing the frequency-dependent noise diode calibration and baseline fitting, we chose the FFT filter approach to remove standing waves \citep[more details can be found in ][]{Xu2025}.
Then we used the ``MedMed'' method to estimate and subtract the off-source spectra, followed by post-processing with the ``poly-asym2'' method.
Finally, after finishing careful RFI flagging, frame conversion, and velocity correction processes, the final calibrated spectra were gridded into a data cube with a pixel size of 1$^\prime$.
The velocity resolution of the data was smoothed from 1.6 km s$^{-1}$ to 3.2 km s$^{-1}$ by a Hanning smooth.
It is worth noting that the data from September 13 contained severe RFI and standing waves, which significantly affected the quality of the final data cube; consequently, these observations were excluded.
The final FAST data cube, with an effective integration time of 225 s per position, has an rms level of 0.9 mJy beam$^{-1}$, which corresponds to a 3 $\sigma$ $\hi$ detectable column density of $\rm 3.9\times10^{17}\,cm^{-2}$ assuming a line width of 10 $\rm km\,s^{-1}$.
For comparison, the rms level of the ALFALFA observation of VCC 693 is 2.13 mJy beam$^{-1}$, corresponding to an $\hi$ column density of $\rm 1.6\times10^{18}\,cm^{-2}$.

\subsection{Optical imaging data}\label{subsec:optical data}
\begin{figure*}[htbp]
	\centering
	\includegraphics[width=0.95\linewidth]{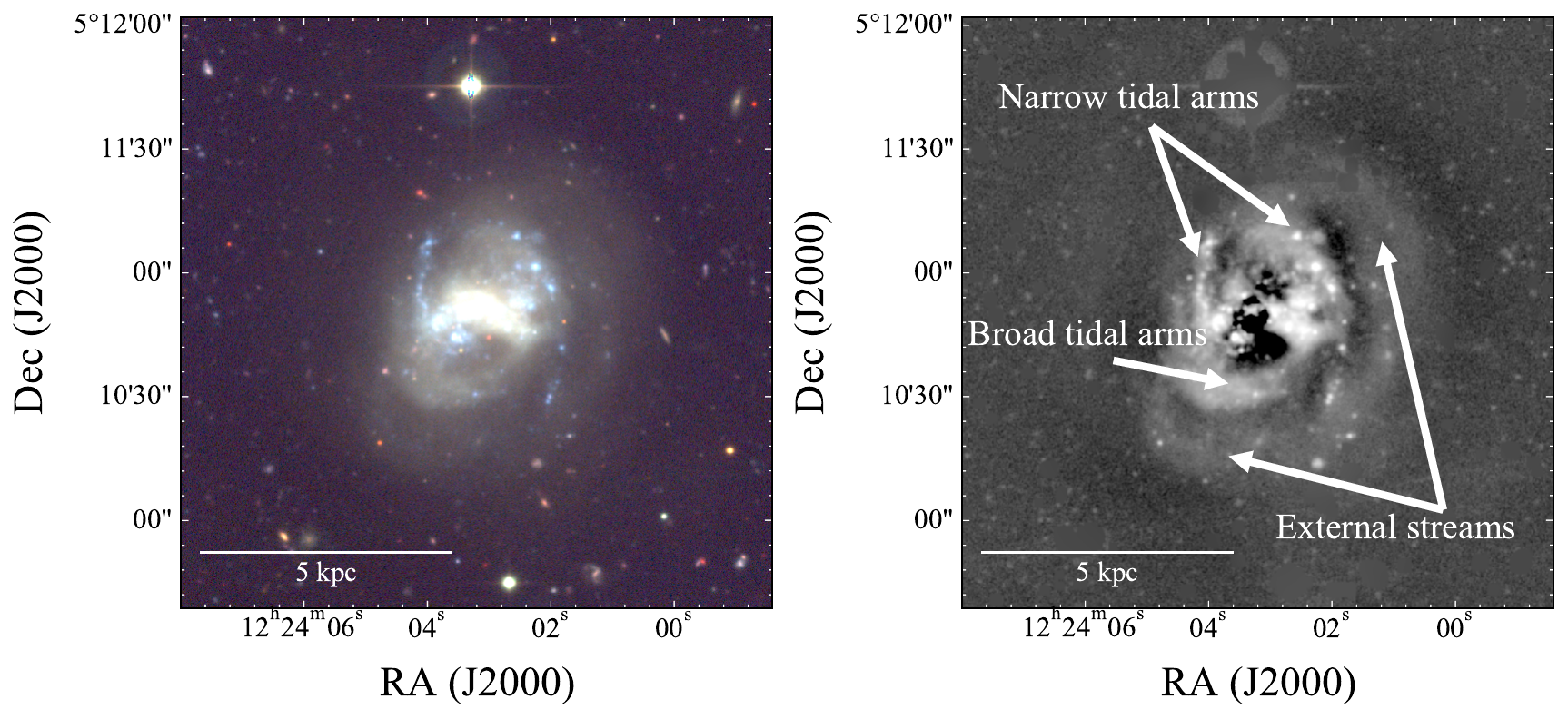}
	\caption{Left panel: the NGVS three-color image of VCC 693, produced by combining the {\em u-}, {\em g-}, and {\em i-}bands. Right panel: the unsharp-masked image of VCC 693.}
	\label{fig:optical_image.pdf}
\end{figure*}

We used broad-band {\em u-}, {\em g-}, {\em i-}, and {\em z-}band images of VCC 693 from the Next Generation Virgo Cluster Survey \citep[NGVS,][]{Ferrarese2012} observed with the MegaCam instrument on the Canada-France-Hawaii Telescope (CFHT).
The image products of NGVS achieved an average 2$\sigma$ surface brightness limit of $\mu_{g}\simeq 29$ mag arcsec$^{-2}$ in {\em g-}band, with a pixel size of $0.^{\prime \prime}187$.
This high sensitivity of the NGVS enables us to detect faint structures and stellar halo components in the outer regions of galaxies.

The narrow-band H$\alpha$ image of VCC 693 was obtained from the Virgo Environmental Survey Tracing Ionised Gas Emission \citep[VESTIGE,][]{Boselli2018}.
The observations were carried out using CFHT and reached a surface brightness sensitivity of $\rm \Sigma(H\alpha)\sim 2\times 10^{-18}\,erg\,s^{-1}\,cm^{-2}\,arcsec^{-2}$ for extended sources when smoothed to $\sim$ 3$^{\prime\prime}$ angular resolution.
The final data were continuum subtracted and corrected for the contamination of $\nii$ lines using the emission line ratio $\nii$/$\rm H\alpha$ from SDSS spectra.

\subsection{SDSS spectra and MUSE data}\label{subsec:MUSE}
We retrieved the single aperture fiber spectra from the SDSS and BOSS spectroscopic surveys observed with the 2.5m Sloan Telescope \citep{Gunn2006,Smee2013}.
The emission line fluxes and stellar kinematics were measured using the Penalized Pixel-Fitting codes \citep[\texttt{PPXF},][]{Cappellari2004}.
\texttt{PPXF} fits a set of weighted stellar population models and Gaussian emission line templates to derive emission line fluxes and stellar kinematics.

In addition to the SDSS spectra, integral field unit spectroscopic observations of VCC 693 obtained with the Multi Unit Spectroscopic Explorer \citep[MUSE,][]{Bacon2014} instrument on the Very Large Telescope (VLT) are available in the archive.
It was observed as part of programme 098.A-0364(A) (PI: Fossati Matteo), covering the southern edge of VCC 693, with the aim of obtaining a comprehensive view of how environmental effects shape the evolution of low-mass systems in clusters.
The MUSE observations of VCC 693 comprise a total of 1937 seconds on-source time, using the MUSE Wide Field Mode with a field of view of $\rm \sim 1\,arcmin^{2}$ and a spatial scale of 0.2 arcsec\,pixel$^{-1}$.
We performed the data reduction using the MUSE pipeline in the \texttt{EsoReflex} environment \citep{Freudling2013}. 
\texttt{EsoReflex} employs a workflow engine that provides visual guidance of the data reduction cascade, including standard processes such as wavelength and flux calibration, sky subtraction, cosmic-ray rejection, and combination of multiple exposures. 
For sky subtraction, we used dedicated 360 seconds of offset sky exposures and further used the Zurich Atmosphere Purge package \citep[\texttt{ZAP};][]{Soto2016} to improve sky subtraction.
We used the Data Analysis Pipeline (\texttt{DAP}) from \citet{Emsellem2022}\footnote{\url{https://gitlab.com/francbelf/ifu-pipeline}} to derive final data products.

The spectral continuum has a S/N too low to be used for stellar population analysis. So we focused  on the nebular emission line properties of star-forming regions visually identified from the H$\alpha$ emission line image (see Section \ref{subsec: metallicity}). 
The spaxels within individual star-forming regions are summed up to increase the spectral S/N. We estimate the oxygen abundance using the O3N2 calibration method of \citet{Pettini2004}, based on a combination of the emission lines $\oiii\ \lambda 5007$, $\nii\ \lambda 6583$, H$\alpha\ \lambda6563$, and H$\beta\ \lambda4861$. The O3N2 method outperforms other widely used methods, such as the $R_{23}$ and N2 methods, at metallicities above $\rm 12+\log (O/H) = 8.1$, with a 1$\sigma$ systematic uncertainty of 0.14 dex.
All emission lines of these star-forming regions used for estimating star formation rate (SFR) and metallicity have a S/N larger than 3 and have been corrected for foreground extinction using the \citet{Fitzpatrick1999} Milky Way extinction curve, assuming $\rm R_{V}=3.1$ and $\rm A(H\alpha)=0.81A_{V}$.
The internal extinction is estimated using the Balmer decrement, assuming an intrinsic $\rm H\alpha/H\beta$=2.86 and adopting the extinction curve of \citet{Calzetti2000}.

\section{Overview of the stellar morphology and environment of VCC 693}\label{sec: stellar properties}

\subsection{Fine stellar morphological features}
In the left panel of Figure \ref{fig:optical_image.pdf}, we present the NGVS three-color composite image of VCC 693.
This galaxy shows a peculiar stellar distribution with multiple morphological features.
The inner region is dominated by a bar structure that exhibits asymmetry around the optical centre and is skewed toward the southwest along the optical minor axis.
Meanwhile, the bar is offset from the $\hi$ kinematic centre of the galaxy, positioned to the north of it (see Section \ref{sec: HI properties} and \ref{sec: HI kinematics}).
Besides the central bar, the main body of VCC 693 appears to be characterised by a series of arms wrapping around the centre, with the inner ones being brighter than the outer ones. 
One of the most prominent broad arms emanates from the east, extends to the south, and then bends toward the northwest. 
There are also two narrow, blue tidal arms, one seemingly emanating from the central region towards the northeast and arcing northwards, and the other forming an almost semi-circle west of the centre. 
Furthermore, several fainter outlying streams can be seen more clearly after applying unsharp-masking to the images (see below).
The plethora of widespread fine structures indicates that the galaxy has been gravitationally perturbed or has experienced a merger event. 

To more clearly bring out this galaxy's finer detail, we applied unsharp masking to the {\em g-}band image, and the resultant image is shown in the right panel of Figure \ref{fig:optical_image.pdf}.
Briefly, the unsharp-masking technology applies elliptical Gaussian smoothing to the original image, with Gaussian sigma smoothly increasing as a function of distance from the centre of the galaxy.
The smoothed image is then subtracted from the original image, and the residual is multiplied by a scale factor of 4 and added to the original image in order to obtain the unsharp masked image \citep[e.g.,][]{Malin1983,McIntosh2008}.
In the unsharp-masked image, the central bar and the above-mentioned fine structures are indicated.
In addition, the barely visible intertwined tidal arms embedded in the halo-like outer regions are brought out clearly in the unsharp masked image. 
There are several blue star-forming clumps associated with the outer tidal arms. 

\subsection{Environment of VCC 693}
\begin{figure*}[htbp]
	\centering
	\includegraphics[width=1\linewidth]{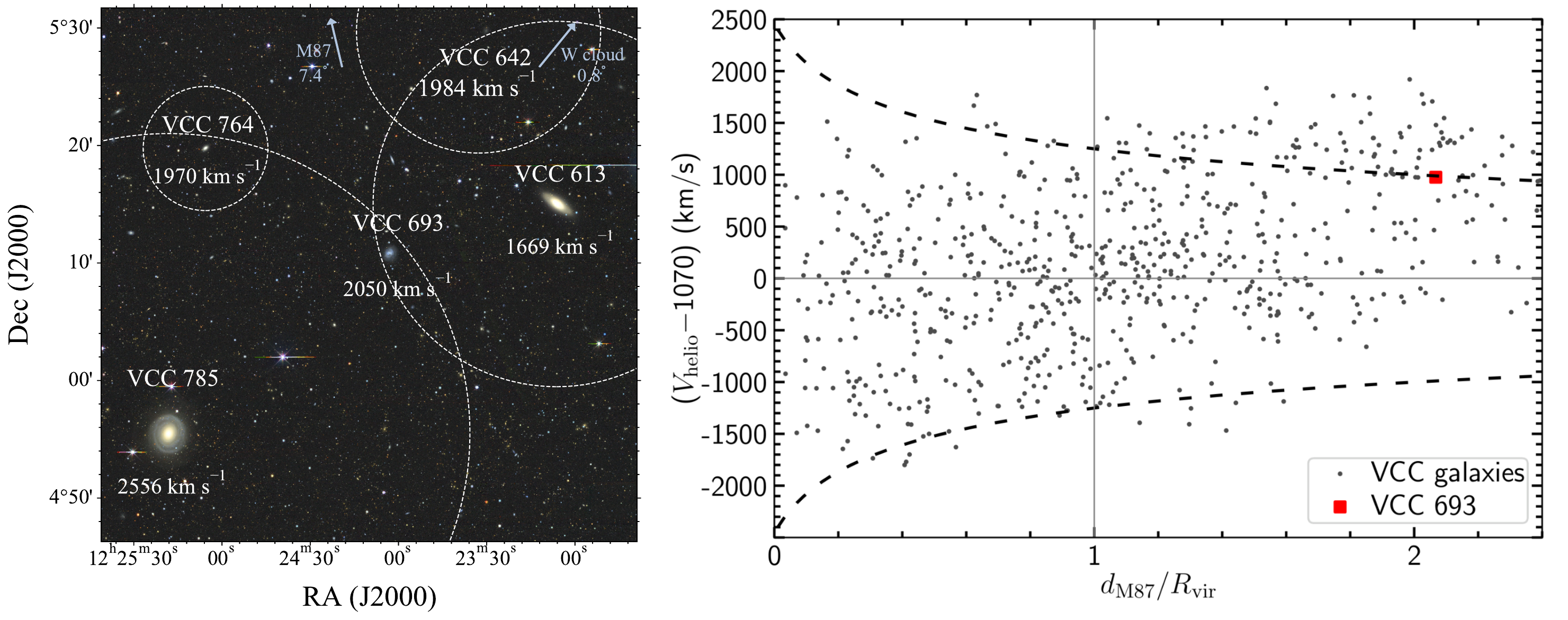}
	\caption{Left panel: The optical image from DESI of the environment of VCC 693. Two galaxies are located to the northwest (VCC 613) and southeast (VCC 785) of VCC 693, whereas a dwarf galaxy is located to the northeast (VCC 764). The dashed lines show their virial radii estimated as 67 times the r-band half-light radius. The blue arrows indicate the directions and angular distances to the centre of the Virgo cluster (M87) and the nearest substructure (W cloud), relative to VCC 693. Right panel: Projected phase-space diagram of redshift-confirmed Virgo galaxies. The escape velocity profiles of Virgo as a function of the angular distance from M87 are indicated by the two black dashed lines. VCC 693 is shown in red, overlaid on the Virgo cluster members displayed in black.}
	\label{fig:VCC693_environment_1.pdf}
\end{figure*}

\begin{figure*}[htbp]
	\centering
	\includegraphics[width=0.95\linewidth]{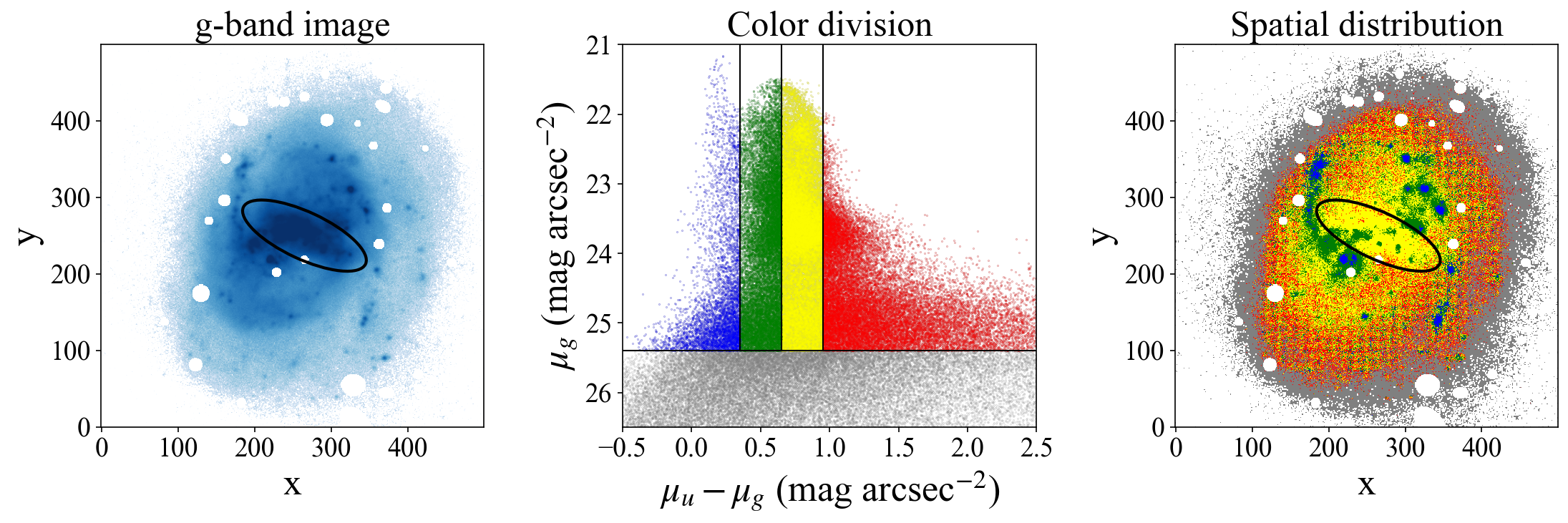}
	\caption{Left panel: NGVS {\em g-}band image after masking background galaxies. Middle panel: Diagram showing {\em u-}band surface brightness versus $u-g$ colors. The analysis is limited to pixels brighter than $\mu_{g}$ = 25.4 mag arcsec$^{-2}$, and the pixels are divided into four tranches according to their $u-g$ colors. The color boundaries between blue, green, yellow, and red tranches are 0.35, 0.65, and 0.95. Right panel: Spatial distribution of the $u-g$ colors, where the colors correspond to those of the tranches in the middle panel. The ellipse in the left and right panels marks the location of the stellar bar.}
	\label{fig:VCC693_u_g.png}
\end{figure*}

The left panel of Figure \ref{fig:VCC693_environment_1.pdf} shows the optical image from the DESI Legacy Imaging Surveys \citep{Dey2019}, displaying the local environment of VCC 693.
VCC 693 lies at the periphery of the Virgo cluster, near its southernmost edge, and is located between two massive galaxies (VCC 785 and VCC 613), both of which have larger stellar masses and contain significant $\hi$ gas reservoirs.
VCC 785 and VCC 613 have projected separations of 116 and 71 kpc, and velocity separations of 506 and $-$381 $\rm km\,s^{-1}$ from VCC 693, respectively.
There are also two dwarf galaxies, VCC 764 to the northeast and VCC 642 to the northwest, with projected separations of 86 and 99 kpc, and velocity separations of $-$80 and $-$66 $\rm km\,s^{-1}$ from VCC 693, respectively.
Their systemic velocities are shown in the figure, and the dashed lines show their virial radii, which are approximately estimated as 67 times their {\em r-}band half-light radii from \citet{Kim2014}.
We present the $\hi$ properties of these neighboring galaxies obtained from our FAST observations and the ALFALFA data in Appendix \ref{sec:neighbor_HIspectra}.
Additionally, based on the NGVS, 11 previously uncatalogued (probable or likely) Virgo members are found within the projected distance from VCC 693 to VCC 785 (24$^{\prime}$), all of which are very faint and lack measured radial velocities.

The right panel of Figure \ref{fig:VCC693_environment_1.pdf} shows the projected phase-space diagram, constructed using the line-of-sight velocity differences between individual galaxies and the average cluster velocity from \citet{Kashibadze2020}, as well as their projected distances from the cluster centre normalized by the virial radius of the Virgo cluster \citep[0.998 Mpc by][]{Simionescu2017}.
The three-dimensional cluster-centric distance of VCC 693 is likely well beyond twice the virial radius of the Virgo cluster. 
Its location in phase space, combined with its gas-normal nature and the near absence of asymmetric $\hi$ stripping features (see Section \ref{sec: HI properties}), suggests that VCC 693 resides in the cluster's periphery and has not yet passed through the virial radius \citep[e.g.,][]{Cen2014}.
As we will demonstrate later through numerical simulations, the rich array of fine stellar tidal features and the regularly rotating $\hi$ disk observed in VCC 693 can be explained by a nearly major merger between a gas-rich and a gas-poor dwarf galaxy (i.e., a damp merger), in contrast to a wet merger or a merger with a gas-free galaxy (which we define as a mixed merger). 
In comparison, optical morphologies resulting from non-merging fly-by tidal interactions are typically dominated by single tidal features, such as symmetric arms. 
Furthermore, as discussed above, since VCC 693 has not yet entered the cluster core region, the occurrence of multiple fly-by interactions \citep[i.e., harassment;][]{Moore1996} required to produce such a diverse set of tidal features is very unlikely.

\section{Quantifying the stellar distributions} \label{sec:stellar and metallicity}
\subsection{Stellar population variation across VCC 693}

\begin{figure*}[htbp]
	\centering
	\includegraphics[width=0.95\linewidth]{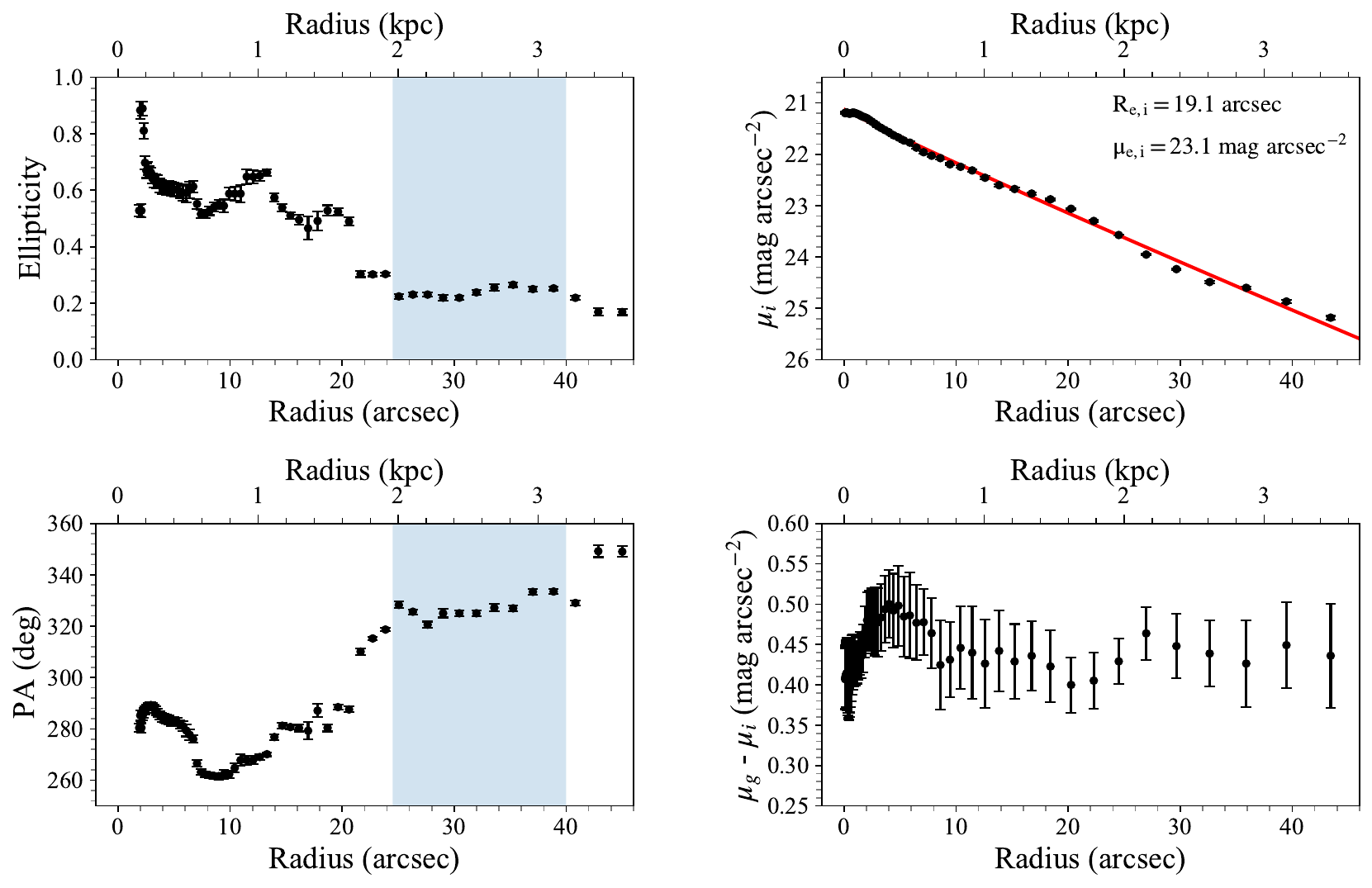}
	\caption{Isophotal analysis of the NGVS images of VCC 693. Left: The radial profiles of the ellipticity (top) and position angle (PA; bottom). Right: The radial profiles of the {\em i-}band surface brightness (top) and $g-i$ color (bottom). The red line shows the \sersic\, profile best fit to the surface brightness profile, with the fitted values of effective radius $R_{\rm e,i}$ and effective surface brightness $\mu_{\rm e,i}$ are shown in the panel. The blue shadow regions in the two left panels mark the region used to derive the disk $\varepsilon$ and PA.}
	\label{fig:optical_isophotal_fit_all.pdf}
\end{figure*}

To probe the spatial distribution of the stellar populations of VCC 693, we analyzed the spatial distribution of different color components \citep[e.g., ][]{Lanyon-Foster2007,Lee2017}.
First, after masking the background objects around the galaxy, we created pixel maps for each band and calculated the surface brightness at each pixel.
The left panel of Figure \ref{fig:VCC693_u_g.png} shows the masked NGVS {\em g-}band pixel map.
The brightest regions of the galaxy are located in the central bar and in sparsely distributed star-forming regions along the narrow tidal arms.
Furthermore, several star-forming regions can also be seen embedded along the outer tidal arm in the {\em g}-band image, with the most significant ones located in the west.

We choose the {\em u}$-${\em g} colors to achieve optimal sensitivity to stellar ages (middle panel of Figure \ref{fig:VCC693_u_g.png}).
The analysis was limited to pixels brighter than $\mu_{g}$ = 25.4 mag arcsec$^{-2}$, which corresponds to 2$\sigma$ surface brightness sensitivity of the {\em g-}band image, and the $u-g$ colors were divided into four tranches.
The boundaries between the blue, green, yellow, and red tranches were determined visually to separate the most prominent spatial components of the stellar distribution and were set at values of 0.35, 0.65, and 0.95, respectively.
The spatial distributions of these four tranches of the $u-g$ colors are shown in the right panel of Figure \ref{fig:VCC693_u_g.png}.
The pixels in the blue tranch are located in several isolated star-forming regions, some of which also exhibit high {\em g-}band brightness, indicating the presence of young stellar populations.
These regions appear to be connected by the green tranch, forming narrow tidal arm structures.
It is worth noting that the long central bar is composed of pixels, nearly all of which are in the yellow tranch, with a redder $u-g$ color than the young stellar population regions and narrow tidal arms, and it also differs from outer regions that are dominated by the red tranch.
The ellipse shown in the left and right panels marks the location of the stellar bar, defined by the sudden jumps in ellipticity and position angle in isophotal analysis. 
The bar length, measured from the isophotal analysis, is $\rm R_{bar}\sim$ 2.62 kpc (see Section 4.2), and it aligns well with the regions traced by the yellow tranch.
The broad tidal arms are dominated by older stellar populations, mainly composed of pixels from the red and yellow tranches.

It is worth noting that the $u-g$ color can also be affected by dust extinction and stellar metallicity, in addition to stellar age. However, the former two factors are expected to have a relatively small effect on the spatial difference of $u-g$ colors compared to stellar ages for low surface brightness late-type dwarfs in general. 
To assess the effect of dust extinction, we have measured the H$\alpha$/H$\beta$ emission-line ratio in the central region and found a value of 3.0, which, assuming an intrinsic ratio of 2.86, implies a dust reddening in $u-g$ of only 0.06 mag---much smaller than the color intervals used to define the different regimes. The reddening effect may be expected to be even weaker in the outer regions. Regarding the effect of metallicity, the central region of VCC 693 has a relatively high metallicity and would therefore be expected to exhibit redder colors at a given stellar age. However, the central bar region of VCC 693 is in fact bluer than the outer, lower-metallicity regions (e.g., the extended tidal arms), unambiguously indicating a younger stellar age in the central region. In addition, there is a good spatial correspondence between the H$\alpha$ emission and the blue-color tranche, further confirming that this tranche hosts the youngest stellar populations.

\subsection{Isophotal analysis of stellar emission}\label{subsec:isophotal}
We conducted isophotal analysis of the optical {\em i-}band images by applying \texttt{photutils} \citep{Bradley2024}.
We performed three iterations to gradually obtain the optical centre coordinates and the radial profiles of ellipticity ($\varepsilon$), position angle (PA), and surface brightness density ($\mu$).
In the first iteration, we kept all of the geometric parameters free and set reasonable initial values for each.
The centre coordinates of the optical galaxy were determined by averaging the converged fitted parameters at large radii and were fixed in subsequent iterations.
Then the radial behaviour of $\varepsilon$ and PA were derived in the second round of fitting, and the results are shown in the left two panels of Figure \ref{fig:optical_isophotal_fit_all.pdf}.
In a barred galaxy, the radial behaviour of $\varepsilon$ often exhibits a sudden decrease as it enters the disk region, whereas the radial profile of PA often displays two distinct values in the bar- and disk-dominated radii.
These patterns are clearly observed in VCC 693 and, due to the offset of the bar, show multiple local peaks or jumps in $\varepsilon$ and PA in the central $\sim1.5$ kpc in radius. To extract the average surface brightness profiles in each band, we used the average $i$-band geometric parameters of $\varepsilon$ and PA of the outer disk where $\varepsilon$ and PA are nearly constant with radius, as indicated by the blue-shaded region in Figure \ref{fig:optical_isophotal_fit_all.pdf}. 

The resulting $i$-band surface brightness radial profile and the $g-i$ color profile are shown in the right panels of Figure \ref{fig:optical_isophotal_fit_all.pdf}. As can be seen, the surface brightness profile largely follows an exponential behaviour, and the color profile become flat ($g-i$ $\sim$ 0.43 mag) beyond the central 10 arcsec. Moreover, we derived the half-light radius $R_{\rm e}$ and the surface brightness $\mu_{\rm e}$ at $R_{\rm e}$ in each band, based on the surface brightness radial profiles, and found that VCC 693 follows the average $R_{\rm e}$-$M_{\rm g}$ and $\mu_{\rm e}$-$M_{\rm g}$ relations of ordinary dwarfs in the Virgo core region \citep{SunW2025}, suggesting that it has overall stellar structural properties indistinguishable from ordinary dwarfs.

To quantify the off-centre stellar bar, we performed a second run of ellipse fitting, using a procedure similar to that described above for the overall stellar light distribution. 
In this case, however, the centre was fixed to the position of the bar rather than to the main body's photometric centre. 
The stellar bar shows an exponential brightness decline from its centre. 
We define the bar length as twice the radius where the ellipticity profile exhibits a sharp drop, with a value of 2.62 kpc, as listed in Table \ref{tab:property}.

\subsection{Current star formation, $\hii$ region metallicity, and velocity field}\label{subsec: metallicity}

\begin{figure*}[htbp]
	\centering
	\includegraphics[width=1\linewidth]{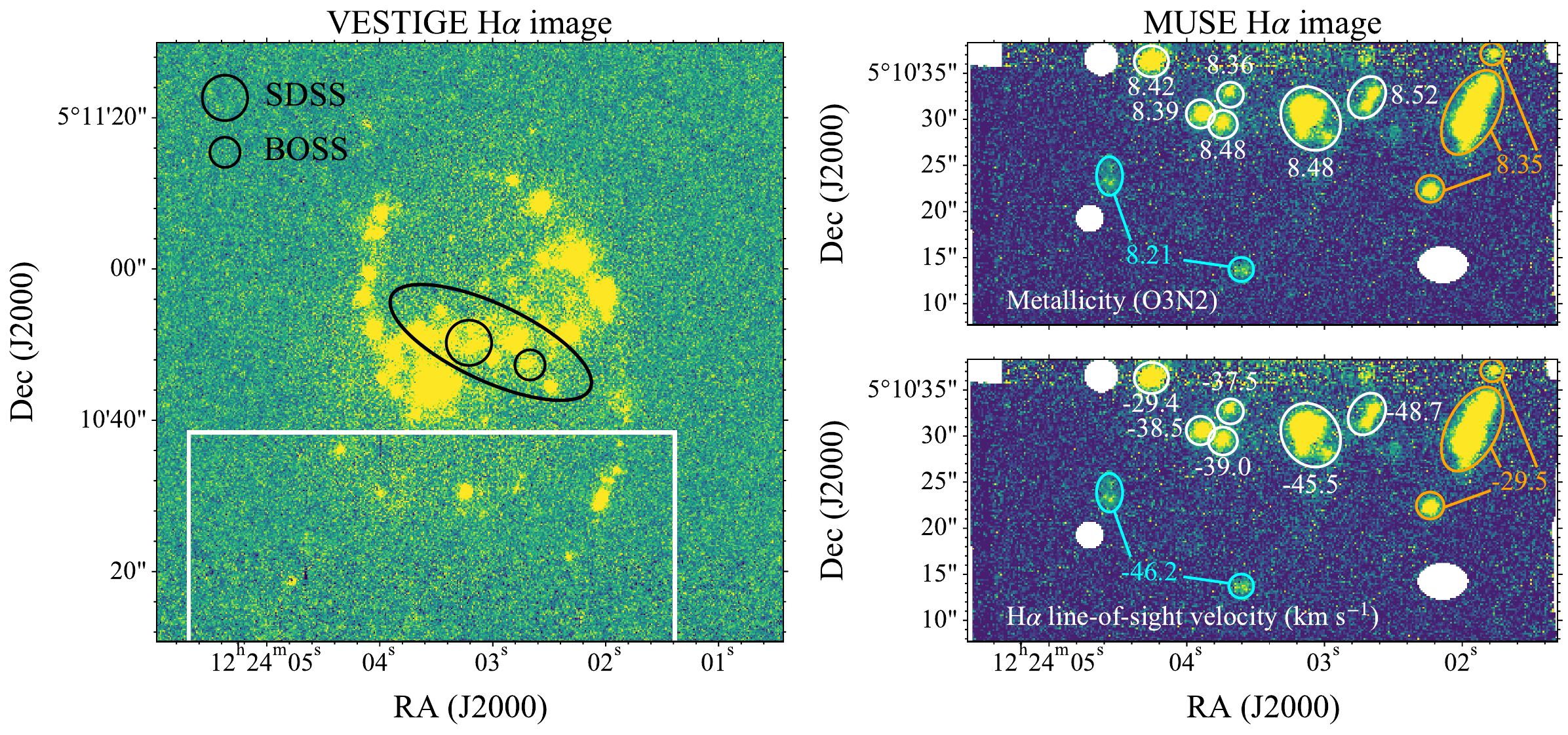}
	\caption{The H$\alpha$ flux density maps from VESTIGE (left) and MUSE (right) observations. The MUSE observations from the archives cover only about a third of the southern region of VCC 693, as shown in the white rectangle. The black circles mark the locations and sizes of SDSS and BOSS spectra fibers. The ellipse in the left panel approximately marks the location of the stellar bar. We identified the star-forming regions with H$\alpha$ emission and measured their metallicity and H$\alpha$ velocity, as indicated in the top right and bottom right panels, respectively. The MUSE star-forming regions are indicated with ellipses of different colors. Spectra of the bottom-left two star-forming regions (in green or blue) and of the rightmost three regions (in yellow) are summed up respectively to increase the S/N. }
	\label{fig:VCC693_MUSE.pdf}
\end{figure*}

\begin{figure*}[htbp]
	\centering
	\includegraphics[width=1\linewidth]{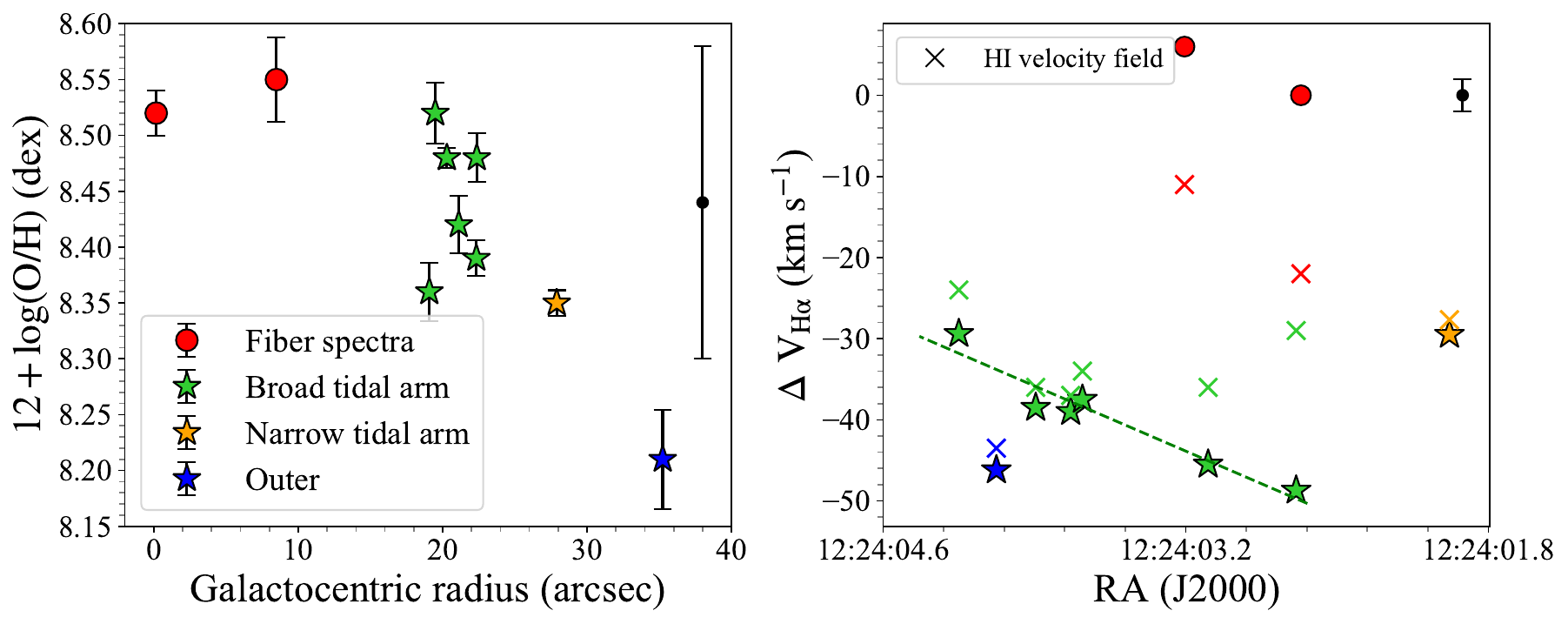}
	\caption{Left panel: the relation between gas-phase metallicity $\rm 12+\log (O/H)$ and galactocentric radius for each star-forming regions. Right panel: the relation between the H$\alpha$ velocity deviations from the systemic velocity and RA. Red, green, yellow, and blue points represent the results from fiber spectra, broad tidal arm, narrow tidal arm, and outer regions, respectively. The $\hi$ velocity deviations from the systemic velocity are indicated by ``x'' in corresponding colors. The error bars of the points in the left panel reflect the measurement uncertainties, whereas the systematic error of the gas-phase metallicity measured by the O3N2 calibration method \citep[$\sim0.14$ dex,][]{Pettini2004}, and the uncertainty caused by the width of the MUSE instrumental profile \citep[$\rm \sim1\,km\,s^{-1}$,][]{Weilbacher2020} are shown in the upper-right corner of each panel. }
	\label{fig:VCC693_SF_region.pdf}
\end{figure*}

We use H$\alpha$ emission as a tracer of recent star formation in VCC 693.
The left panel of Figure \ref{fig:VCC693_MUSE.pdf} shows the continuum-subtracted H$\alpha$ image of VCC 693 from VESTIGE.
The brightest H$\alpha$ emission regions largely overlap with the blue and green tranches in the $u-g$ colors (see Figure \ref{fig:VCC693_u_g.png}).
It reveals widespread H$\alpha$ emission across the galaxy, with the most H$\alpha$-luminous areas found at the galaxy's centre and the narrow tidal arms.
It should be noted that the spatial distribution of H$\alpha$ emission in the inner region largely follows the southeast-northwest direction, which is different from that of the bar marked by the black ellipse.
This aligns with the relatively red optical colors of the bar revealed in Figure \ref{fig:VCC693_u_g.png}.
The locations and sizes of the SDSS and BOSS spectra fibers (3$^{\prime \prime}$ for SDSS and 2$^{\prime \prime}$ for BOSS) are marked in the left panel of Figure \ref{fig:VCC693_MUSE.pdf} with black circles.
The spectra show that VCC 693 lies in the star-formation region of the BPT diagrams \citep{Kauffmann2003}, with no evidence for an AGN or shock features detected.
We estimated the oxygen abundance of the SDSS, BOSS, and MUSE spectra by applying the O3N2 method of \citet{Pettini2004} to extinction-corrected emission lines.
The resultant gas-phase metallicity is $\rm 12+\log (O/H) = 8.52$ for the SDSS spectrum and $\rm 12+\log (O/H) = 8.55$ for the BOSS spectrum.
Considering that VCC 693 has a stellar mass of $\log (M_{\star}/M_{\odot}) = 8.44$, its gas-phase metallicity is $\sim$ 0.2 dex higher than expected from the mass-metallicity relation for $\hi$-bearing dwarf galaxies in \citet{Jimmy2015}, where the same O3N2 calibration method was adopted.
This may suggest enhanced metal enrichment induced by merger-triggered star formation.

The MUSE H$\alpha$ images are shown in the two right panels of Figure \ref{fig:VCC693_MUSE.pdf}, with the MUSE field of view shown by the white rectangle in the left panel.
We identified star-forming regions from the MUSE H$\alpha$ image (marked by ellipses overlaid on the image) and extracted spectra for each region.
The star-forming regions located in the broad tidal arms are marked by white ellipses, while the regions in orange belong to the narrow tidal arms.
The blue ellipses mark two weak star-forming regions embedded in the stellar halo, which are barely visible in the VESTIGE H$\alpha$ image due to its low surface brightness.
We extracted the spectrum for each region enclosed by the white ellipses, whereas we extracted a single combined spectrum, respectively, for the blue and orange elliptical regions to improve the S/N.

We determined the gas-phase metallicity for the above marked regions. 
The gas-phase metallicity exhibits a large gradient between the centre and the outer star-forming regions.
The broad tidal arms also show a wide metallicity range from 8.36 to 8.52.
In the left panel of Figure \ref{fig:VCC693_SF_region.pdf}, we show the relation between gas-phase metallicity $\rm 12+\log (O/H)$ and galactocentric radius for each star-forming region to illustrate the metallicity variation.
These star-forming regions exhibit a negative gradient of $\rm \sim -0.11\,dex/kpc$ ($\rm -0.15\,dex/R_{e}$), with the metallicity in the outer regions dropping to 8.21.
If we exclude the outer, slightly low-S/N star-forming regions and measure only up to the narrow tidal arm, the metallicity gradient remains large at $\rm \sim -0.08\,dex/kpc$ ($\rm -0.11\,dex/R_{e}$), which is at the lower end but significantly steeper than the average for dwarf galaxies with comparable mass \citep{Belfiore2017,Poetrodjojo2021,Li2025}.
This metallicity gradient may not only reflect the natural inside-out growth of the galaxy, but could also be a result of its complex environmental effects or merger events, which have led to disordered $\hii$ regions embedded in disturbed stellar populations.
The high central gas-phase metallicity suggests that its ISM has been enriched by past supernova explosions.
The wide metallicity range in the broad tidal arms indicates their complex formation history, likely driven by tidal interactions, and suggests an inhomogeneous mixture of metal-rich gas and metal-poor gas.
The very low metallicity in the outskirts could be attributed to recent gas inflow of metal-poor gas or a merger event with a low metallicity galaxy.

The H$\alpha$ radial velocities of these star-forming regions are indicated in the bottom right panel of Figure \ref{fig:VCC693_MUSE.pdf}.
The right panel of Figure \ref{fig:VCC693_SF_region.pdf} shows the relationship between the H$\alpha$ and $\hi$ velocity (indicated by ``x''; see Section \ref{sec: HI properties}) deviations from the systemic velocity and RA.
The radial velocities of most star-forming regions broadly agree with the $\hi$ velocities within the uncertainties. 
This general agreement between the ionized gas velocities (as traced by the star-forming regions) and the $\hi$ velocities in narrow tidal arms and outer region suggests that the two components are likely aligned in spatial distribution.
However, the $\hi$ velocity remains roughly constant along the broad tidal arms, whereas H$\alpha$ shows a velocity gradient.
The kinematic discrepancy between the $\hi$ and H$\alpha$ gas in this region suggests a decoupling between the ionized and neutral gas, likely reflecting localized perturbations or recent gravitational interaction.
The velocity offsets observed in the central SDSS/BOSS fiber are likely caused by the dynamical interactions associated with the bar.

\section{$\hi$ gas distribution of VCC 693}\label{sec: HI properties}

\begin{table*}
\caption{$\hi$ properties of VCC 693.\label{tab:HI property}}
\centering
\renewcommand\arraystretch{1.5}
\begin{tabular}{lccccccc}
\hline\hline
Telescope & Resolution & $\rm \sigma_{rms}$ & N($\hi$) (3$\sigma$, 10 $\rm km\,s^{-1}$) & $\rm V_{sys}$& $W_{50}$ & $\rm S_{int}$ & $\rm \log M_{\hi}$ \\
 & & (mJy beam$^{-1}$)& (cm$^{-2}$)& ($\rm km\,s^{-1}$)& ($\rm km\,s^{-1}$) & (Jy $\rm km\,s^{-1}$) & ($M_{\odot}$) \\
\hline
VLA (natural) & $13.^{\prime\prime}4\times9.^{\prime\prime}9$ & 0.52 & $\rm 7.5\times10^{19}$ & 2050.7$^{+0.6}_{-0.4}$ & 99.1$^{+1.3}_{-0.8}$ & 2.40$^{+0.01}_{-0.02}$ & 8.19$^{+0.06}_{-0.06}$\\
VLA (robust) & $7.^{\prime\prime}9\times6.^{\prime\prime}4$ & 0.59 & $\rm 2.2\times10^{20}$ & 2049.3$^{+0.5}_{-0.4}$ & 97.4$^{+0.7}_{-0.9}$ & 2.47$^{+0.02}_{-0.02}$ & 8.20$^{+0.06}_{-0.06}$\\
FAST& $2.^{\prime}9\times2.^{\prime}9$& 0.9 & $\rm 3.9\times10^{17}$ & 2049.7$^{+0.5}_{-0.3}$ & 102.5$^{+1.0}_{-0.7}$ & 2.52$^{+0.01}_{-0.01}$ & 8.21$^{+0.06}_{-0.06}$\\
ALFALFA& $3.^{\prime}8\times3.^{\prime}3$& 2.13 & $\rm 1.6\times10^{18}$ & 2052.9$^{+0.5}_{-0.6}$ & 97.6$^{+1.2}_{-1.2}$ & 2.74$^{+0.03}_{-0.03}$ & 8.24$^{+0.06}_{-0.06}$\\
\hline
\end{tabular}
\end{table*}
 
\subsection{Overview of the $\hi$ distribution}
\begin{figure}[htbp]
	\centering
	\includegraphics[width=1\linewidth]{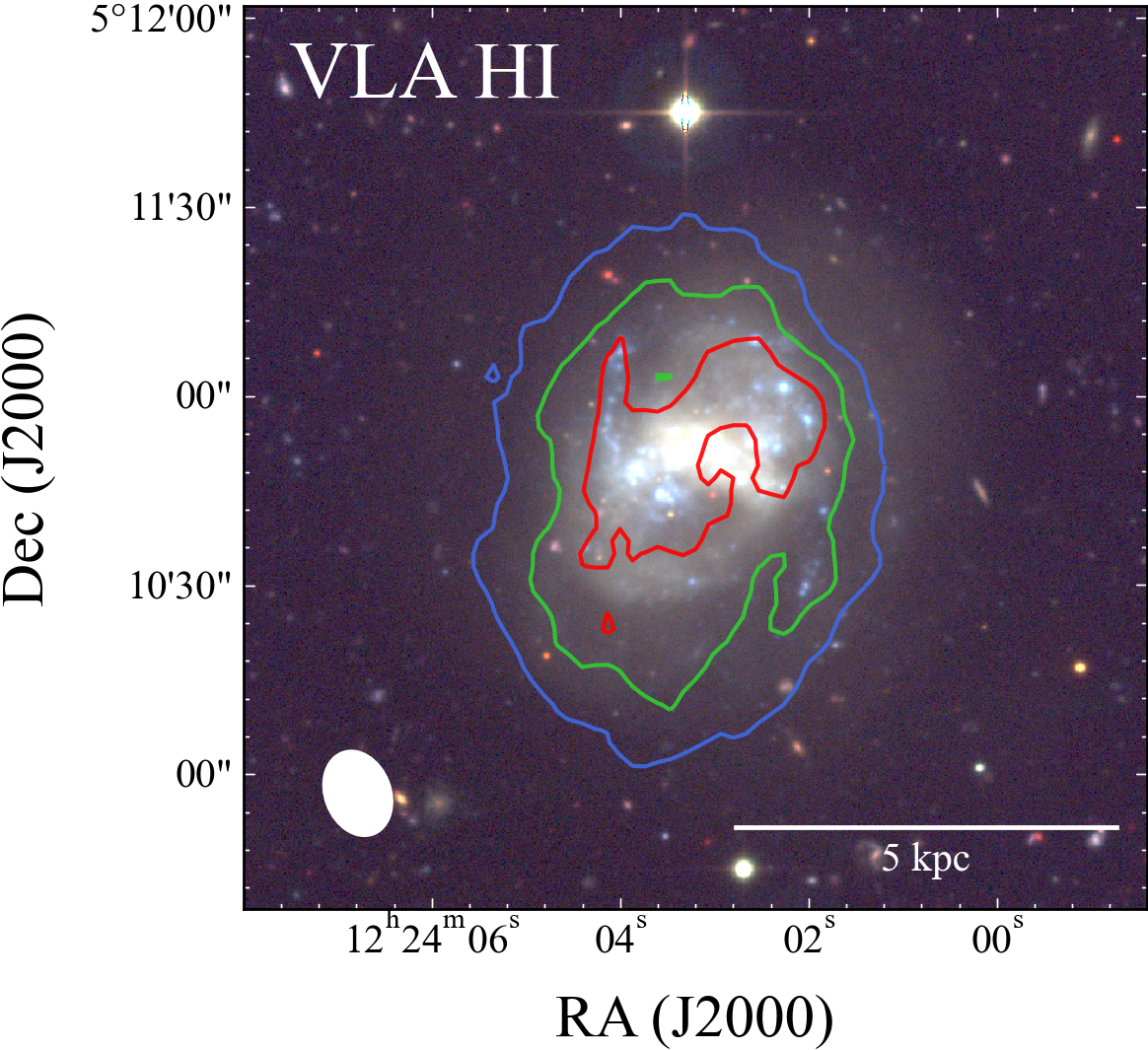}
	\caption{The NGVS three-color image of VCC 693 with the VLA natural-weighted $\hi$ contours overlaid. The blue, green, and red contours are at levels of (2, 6, 10) $\rm \times 10^{20}\,cm^{-2}$. The beam size is shown in the bottom left. }
	\label{fig:VCC693_optical_HI.png}
\end{figure}

\begin{figure}[htbp]
	\centering
	\includegraphics[width=1\linewidth]{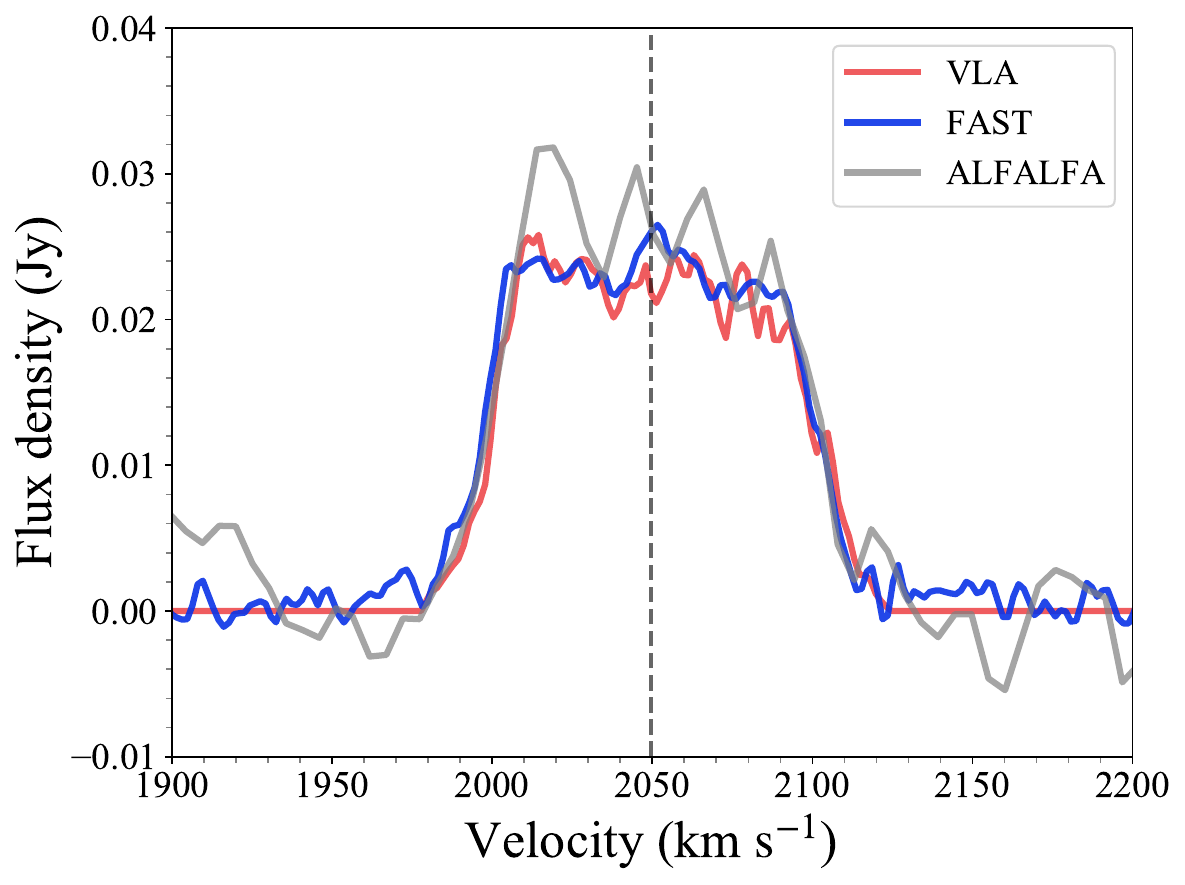}
	\caption{The $\hi$ profile of VCC 693 from our VLA observations (red), our FAST observations (blue), and ALFALFA (grey). The black vertical line shows the $\hi$ systemic velocity of 2049.7 $\rm km\,s^{-1}$ from FAST. }
	\label{fig:HI_spectra.pdf}
\end{figure}

In Figure \ref{fig:VCC693_optical_HI.png}, we present the $\hi$ column density distribution (natural-weighted {\sc moment0} map) overlaid as contours on the three-color NGVS image of VCC 693.
At the detection limit of $\rm 2\times 10^{20}\,cm^{-2}$ of our VLA observations, the $\hi$ gas spans a region similar to that of the stellar components.
The $\hi$ distribution is largely symmetric and lacks evidence of environmental stripping, except for a slight compression of the column density contours at the northwest edge relative to the southeast, and a lesser extent than the stellar tidal features.
The highest $\hi$ surface density is found in regions of narrow tidal arms with young stellar populations, whereas the bar region has relatively lower $\hi$ surface density.
We estimated an $\hi$ radius of 44 arcsec at the N($\hi$) = 1 $\rm M_{\odot}\,pc^{-2}$ contour level along the major axis of the natural-weighted $\hi$ map.
Compared to the $\rm R_{25}\sim 29\,arcsec$ (the semi-major axis of the 25 mag arcsec$^{-2}$ isophote) from the optical $B$-band image \citep{Paturel1991}, the resulting $\hi$-to-optical size ratio $\rm R_{\hi}/R_{25}$ is $\sim1.5$, which is slightly lower than the typical values of $\sim1.7\pm0.5$ for normal late-type galaxies \citep{Broeils1997}, suggesting that the perturbation induced by the dynamical interaction of the galaxy with its surrounding environment (hot gas) if present, is minor.

The integrated $\hi$ velocity profiles obtained from VLA, FAST, and ALFALFA are shown in Figure \ref{fig:HI_spectra.pdf}.
The VLA $\hi$ profile is derived from the natural-weighted data cube and shows reasonable agreement with that obtained from FAST.
The profiles exhibit a nearly flat-topped shape with the $\hi$ flux slightly higher toward the lower-velocity end, which spatially corresponds to the southern direction.
From the global $\hi$ profile, we measured the systemic velocity ($\rm V_{sys}$), the velocity width at 50\% ($w_{50}$) of the peak flux, total integrated flux ($\rm S_{int}$), and $\hi$ mass ($\rm M_{\hi}$).
These measurements are given in Table \ref{tab:HI property}.
The uncertainties of these parameters are estimated by repeating the measurements 1000 times on profiles randomly perturbed according to the observational uncertainties.
An additional 10\% systematic uncertainty, as recommended by \citet{Haynes2018} is also included.

\subsection{$\hi$ moment maps}\label{subsec:moment map}
\begin{figure*}[htbp]
	\centering
	\includegraphics[width=0.95\linewidth]{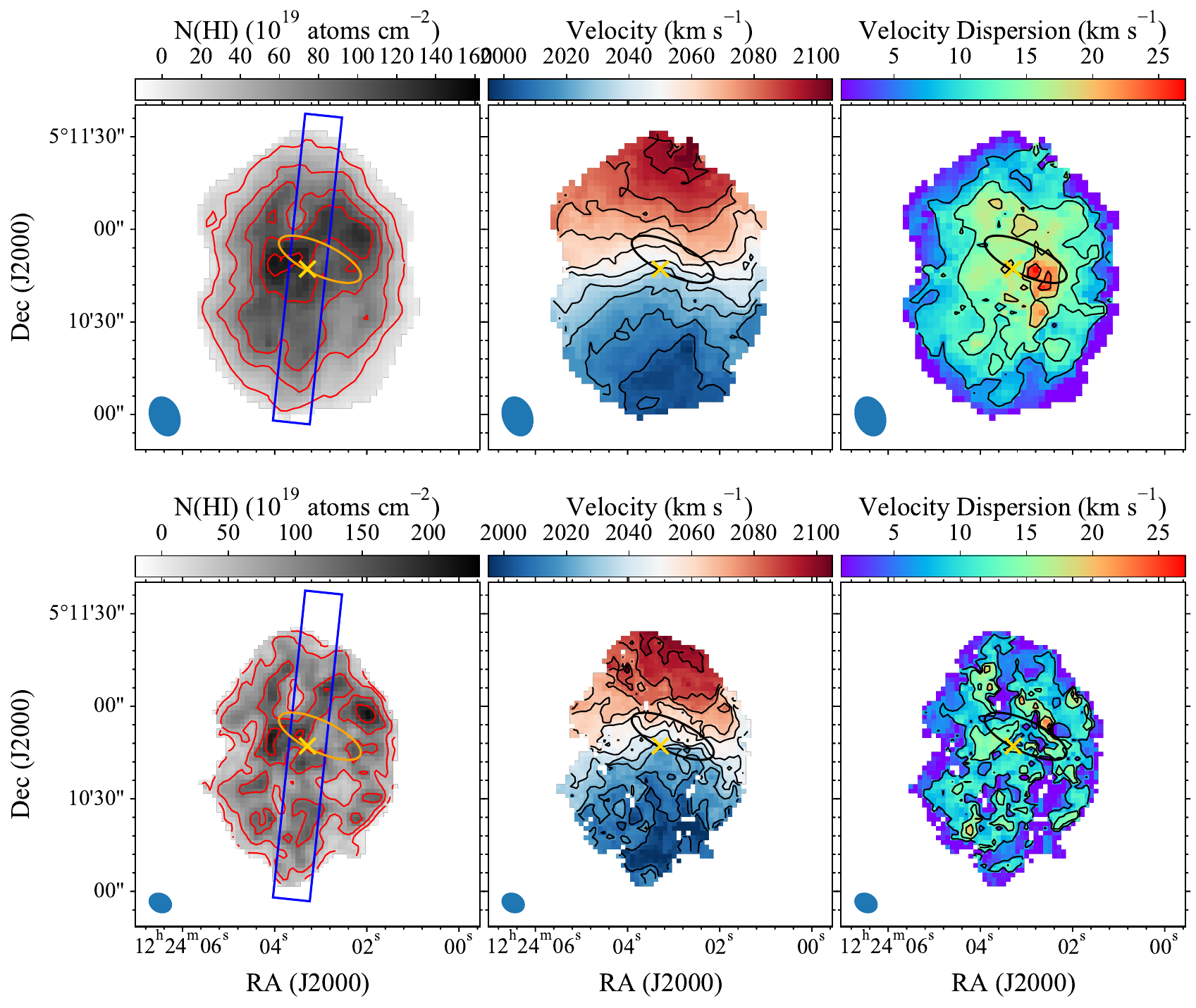}
	\caption{$\hi$ moment maps of VCC 693. The top three panels show the maps calculated from the natural-weighted cube, whereas the bottom three are from the robust-weighted cube. The left panels show the $\hi$ column density maps ({\sc moment0}). The column density contours are at levels of (2, 5, 8, 11, 14) $\rm \times 10^{20}\,cm^{-2}$ for the natural-weighted map and (2, 6, 12, 18) $\rm \times 10^{20}\,cm^{-2}$ for the robust-weighted map. The blue rectangle corresponds to the cut through the galaxy with a width of 12$^{\prime \prime}$, which is used to measure the $\hi$ column density profiles in Figure \ref{fig:VCC693_asymmetry.pdf}. The middle panels show the intensity-weighted $\hi$ velocity field ({\sc moment1}). The velocity contours span the range 2005-2095 $\rm km\,s^{-1}$ with 10 $\rm km\,s^{-1}$ intervals. The right panels show the velocity dispersion maps ({\sc moment2}) and the contours are drawn at 4 $\rm km\,s^{-1}$ intervals. The ellipses and yellow crosses in each panel mark the bar and $\hi$ centre, respectively. The beam sizes of natural- and robust-weighted cubes are shown in the bottom left corner.}
	\label{fig:VLA_moment_map.pdf}
\end{figure*}

\begin{figure}[htbp]
	\centering
	\includegraphics[width=1\linewidth]{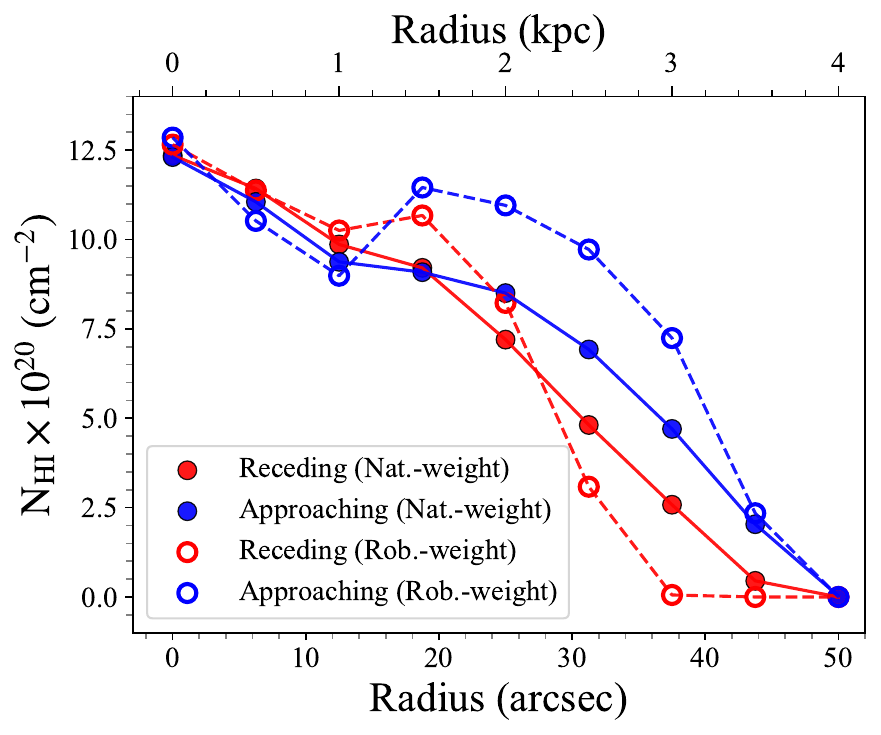}
	\caption{The $\hi$ column density cut profile along the $\hi$ kinematical major axis from the centre in the receding (red) and approaching (blue) sides with a width of 12$^{\prime \prime}$. The solid points shows the results from natural-weighted data cube and the hollow points shows the robust-weighted results. }
	\label{fig:VCC693_asymmetry.pdf}
\end{figure}

The channel maps of the natural-weighted $\hi$ image cube, covering velocities from 1967.7 to 2118.1 km $\rm s^{-1}$ and re-gridded to a channel width of 10 km $\rm s^{-1}$, are presented in Appendix \ref{sec:channel map}.
Figure \ref{fig:VLA_moment_map.pdf} shows the $\hi$ integrated intensity maps ({\sc moment0}), intensity-weighted $\hi$ velocity fields ({\sc moment1}), and velocity dispersion maps ({\sc moment2}) of VCC 693, derived from both the VLA natural-weighted data cube (top) and robust-weighted data cube (bottom).
The $\hi$ kinematic centre (yellow crosses) derived from velocity field fitting (see below) is offset toward the southeast from the optical centre, which approximately lies on the southern edge of the stellar bar.
To quantify the slight asymmetry along the northwest-southeast direction described above, we extracted the average $\hi$ column density profiles along the kinematic major axis direction (see Section \ref{sec: HI kinematics}), which align approximately with the direction of asymmetry. 
The rectangular slit used to extract the column density cuts is 12$^{\prime \prime}$ wide and is indicated in the left two panels in Figure \ref{fig:VLA_moment_map.pdf}.
Figure \ref{fig:VCC693_asymmetry.pdf} shows the column density cut profiles for both the receding (north) and approaching (south) sides.
The receding and approaching sides in the inner regions of the galaxy show similar profiles, with their local peaks and troughs aligned.
The asymmetry appears after $\sim$ 20 arcsec (1.5 kpc), with the receding side exhibiting an overall steeper radial decline, which is more remarkable in the robust-weighted map.
The asymmetry may result from a merger event or mild environmental effect, such as RPS. 
However, it is worth mentioning that $\hi$ profile asymmetry is not unusual even for field or isolated galaxies \citep[e.g.,][]{Espada2011}. 

From the $\hi$ {\sc moment1} maps, the overall $\hi$ velocity gradient is observed over $\sim110$ $\rm km\,s^{-1}$ in the data cubes, with a slight offset relative to the major axis of the optical image.
The $\hi$ kinematics exhibits a regularly rotating disk; however, there is a Z-shaped twist in the central region, which aligns with the optical bar structure (Figure \ref{fig:VCC693_u_g.png}).
Such a Z-shaped twist is commonly observed in barred galaxies and reflects significant gas streaming motions along the bar.
Additional outer twists of isovelocity contours are also found around the minor axis, which may be attributed to warps or radial gas flows. 

The $\hi$ {\sc moment2} maps show that the overall velocity dispersion decreases with radius.
In the natural-weighted map, the largest velocity dispersion is found to the west (and slightly to the southwest) of the $\hi$ centre, near the region of the largest velocity gradient, which may be associated with the bar-induced streaming motion mentioned above. 
However, this high velocity dispersion feature is absent in the robust-weighted map, suggesting that the apparent enhancement in the lower-resolution, natural-weighted map is more likely an artifact of beam smearing. 
In support of this interpretation, the region exhibits a highly skewed line profile.

\subsection{Radial profiles of mass surface density}\label{subsec:radial profiles}
\begin{figure}[htbp]
	\centering
	\includegraphics[width=1\linewidth]{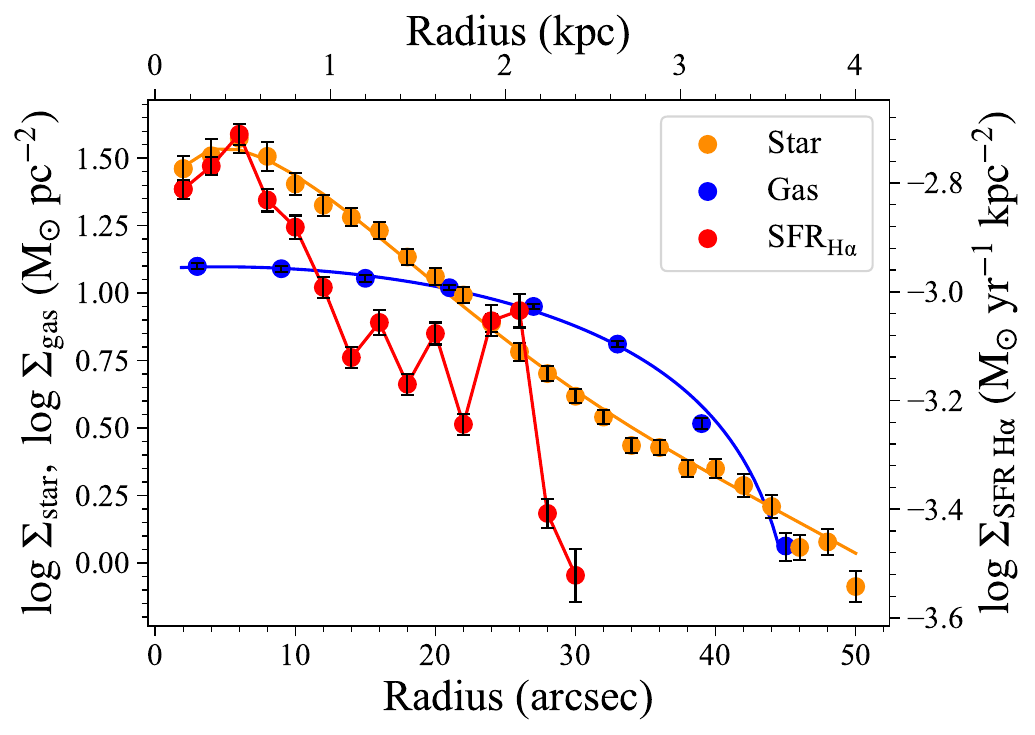}
	\caption{Radial profiles of $\rm \log \Sigma_{\star}$ (orange), $\rm \log \Sigma_{gas}$ (blue), and $\rm \log \Sigma_{SFR}$ (red). The $\rm \Sigma_{\star}$ and $\rm \Sigma_{gas}$ are in units of $M_{\odot}$ kpc$^{-2}$ as indicated on the left label, whereas the $\rm \Sigma_{SFR}$ is in units of $M_{\odot}$ yr$^{-1}$ kpc$^{-2}$ as indicated on the right label. The lines of gas and stellar components show the best-fit profiles.}
	\label{fig:VCC693_density_profile.pdf}
\end{figure}
Before moving to the next section about mass profile decomposition of VCC 693, here we derived the radial profiles of the stellar mass, atomic gas mass, and SFR surface density using the $\hi$ kinematic parameters: the kinematic centre, position angle, and inclination (see Section \ref{sec: HI kinematics}).
We performed azimuthal averaging in concentric elliptical rings with a width of 2$^{\prime \prime}$ for the stellar and SFR surface densities, and 6$^{\prime \prime}$ (almost half the beam size) for $\hi$.
The mean value within each ring is then plotted as a function of radial distance from the kinematic centre.

The stellar mass surface density profile was derived from the {\em i-}band surface brightness based on the $g-i$ color-dependent mass-to-light ratio calibrated for Local Group dwarf galaxies from \citet{Zhang2017}, assuming a \citet{Chabrier2003} initial mass function: $\log(M_{\star}/L_{\rm i}\,[M_{\odot}/L_{\odot,i}])\rm =-0.471+0.720\times(g-i)$.
For the atomic gas mass surface density profile, we used the natural-weighted $\hi$ surface density map and applied a factor of 1.36 to account for the presence of Helium and metals.
The SFR surface density was estimated from the VESTIGE H$\alpha$ map using the relation: $\rm \log(SFR_{H\alpha}[M_{\odot }\, yr^{-1}])=\log(5.5\times10^{-42}\, L_{H\alpha}\, [erg\,s^{-1}])$, where $L_{\rm H_{\alpha}}$ is the $\rm H_{\alpha}$ luminosity corrected for the contamination of $\nii$ lines and extinction (see Section \ref{subsec:optical data} and \ref{subsec:MUSE} for details).

Figure \ref{fig:VCC693_density_profile.pdf} presents the radial profiles.
The stellar and SFR surface density radial profiles are fit with a `poly-exponential' disk, as shown by the lines.
As commonly observed in normal galaxies, the atomic gas surface density profile of VCC 693 shows a relatively flat central region, followed by a steep radial decline beyond approximately 2.5 kpc.
The stellar mass surface density profile broadly follows an exponential decline across the main body of the galaxy. 
In contrast, the overall SFR surface density profile exhibits a significantly steeper radial drop---apart from several bumps associated with $\hii$ regions in the outer tidal arms---than the stellar mass surface density profile. 
Such a steep decline in the SFR surface density is atypical for normal galaxies with comparable baryonic mass \citep[$\sim4.3\times10^{8}\,\rm M_{\odot}$;][]{Zhang2012}.

\section{$\hi$ kinematics of VCC 693}\label{sec: HI kinematics}

\subsection{Kinematic modelling}\label{subsec:barolo}
By fitting either 2D or 3D tilted-ring models, a galaxy's rotation curve can be reconstructed from $\hi$ data cubes. 
In the 2D tilted-ring approach, the galactic disk is represented as a set of concentric elliptical rings, each characterized by a kinematic centre, systemic velocity, position angle, inclination, radial expansion velocity, and rotational velocity.
The model is fitted to a 2D velocity field extracted from the $\hi$ data cube, which is typically constructed from representative centroid velocities of the line-of-sight profiles, such as {\sc moment1} maps, single-Gaussian fitting, or higher-order Gauss-Hermite h3 polynomial fitting \citep[e.g.,][]{Oh2018,Oh2019}.

In contrast, 3D tilted-ring models constrain galaxy kinematics by fitting directly to the full $\hi$ data cube \citep[e.g.,][]{Jozsa2007,DiTeodoro2015}. 
In this approach, a synthetic cube is generated by modelling the gas disk as a set of concentric rings with finite thickness, populated with emitting gas clouds that follow prescribed rotational and velocity-dispersion profiles. 
The model cubes are convolved to match the telescope beam and spectral resolution, and then compared to the observed cube on a channel-by-channel basis. 
By accounting for beam-smearing effects and mitigating biases that could be caused by the manner of velocity-field extraction, the 3D approach provides more reliable rotation curves, particularly for low-resolution data or highly inclined galaxies. 

An observed line-of-sight velocity $V_{\rm los}$ consists of three components: 
\begin{eqnarray}
V_{\rm{los}} = V_{\rm{sys}}+(V_{\rm{rot}}\,\rm{cos}(\theta) + V_{\rm{rad}}\,\rm{sin}(\theta))\rm{sin}(\it i)
\end{eqnarray}
where $V_{\rm sys}$, $V_{\rm rot}$, and $V_{\rm rad}$ are the systemic velocity, rotation velocity component, and radial flow velocity component, respectively.
Here $i$ is the inclination angle and $\theta$ is the azimuthal angle in the plane of the disk.
After subtracting $V_{\rm sys}$, the $V_{\rm los}$ in gas-rich galaxies is typically dominated by rotation. 
The radial flow velocity component, if any, is most obvious near the minor axis. 
We used $\rm ^{3D}$Barolo on the natural-weighted data cube, adopting a ring separation of 6$^{\prime \prime}$ (half the beam size). 
$\rm ^{3D}$Barolo associates a series of parameters with the observed $V_{\rm los}$ in order to model the spectroscopic data cubes and derive the best-fit parameters for each individual ring. 
To break the degeneracy between radial gas flows and gas disk warps, we followed the multi-step technique from \citet{DiTeodoro2021}.
We divided the kinematic modelling process into three stages, involving a standardized procedure to derive a rotation curve and a further step to determine any contribution from radial gas flows:

(i) Pre-run.
At this stage of fitting, our aim is to derive the centre and systemic velocity of the galaxy.
During this fitting process, all parameters were kept free, and the initial parameters for the galaxy centre ($x_{0}$, $y_{0}$), inclination angle ($i$), and position angle ($\Phi$) were set to that determined from the optical image, whereas the systemic velocity was obtained from the midpoint of the global $\hi$ profiles.
We inspected the fitting results visually to ensure their convergence.

(ii) First stage.
This stage aims to determine the rotation curve and potential disk warps.
We follow a classical modelling procedure that includes two rounds of fitting.
The first round allows for radial variations of $i$, $\Phi$, $V_{\rm rot}$, and $\sigma_{\rm gas}$.
After regularizing $i$ and $\Phi$ to a constant value and a Bezier curve, respectively, a second round of fitting was performed with only the $V_{\rm rot}$ and $\sigma_{\rm gas}$ parameters kept free.
In these two rounds of fitting, $V_{\rm rad}$ was set to 0, and we used an azimuthal weighting of the data of $(\rm{cos}\,\theta)^{2}$.
Additionally, pixels within 20$^\circ$ of the minor axis were not taken into consideration in order to minimize the influence of potential radial motions.

(iii) Second stage.
At this stage, we aim to determine $V_{\rm rad}$ while keeping all other parameters fixed.
Radial motions primarily affect regions near the kinematic minor axis; therefore, we applied a $(\rm{sin}\,\theta)^{2}$ weighting during the fitting process and excluded pixels within 20 degrees of the major axis, where radial velocity components are most prominent.

The errors were estimated with $\rm ^{3D}$Barolo's default error calculation algorithm via the Monte Carlo method.
The first and second runs were repeated three times to determine the parameters for the approaching side, the receding side, and both sides of VCC 693.
A reasonable mask that can effectively reduce the influence of noise is important during the kinematics fitting, so we tested two different mask methods.
One mask was built by $\rm ^{3D}$Barolo using the source-finder algorithm, with primary and secondary thresholds of $5\times \sigma_{\rm rms}$ and $2\times \sigma_{\rm rms}$, respectively.
The other was obtained with the \texttt{SoFiA-2} source finding software.
We adopted the second masking method, as using the first one showed no significant differences in the radial trends of parameters.
Lastly, to obtain the circular velocity $V_{\rm circ}$, we corrected $V_{\rm rot}$ for the effect of pressure support by applying an asymmetric drift correction to the fitted rotation curves \citep{Iorio2017}.
The circular velocity provides a more direct measure of the gravitational potential and dynamical mass of a galaxy, with the correction increasing the rotation curve by $\sim$3 $\rm km\,s^{-1}$ on average for VCC 693.

\begin{figure*}[htbp]
	\centering
	\includegraphics[width=1\linewidth]{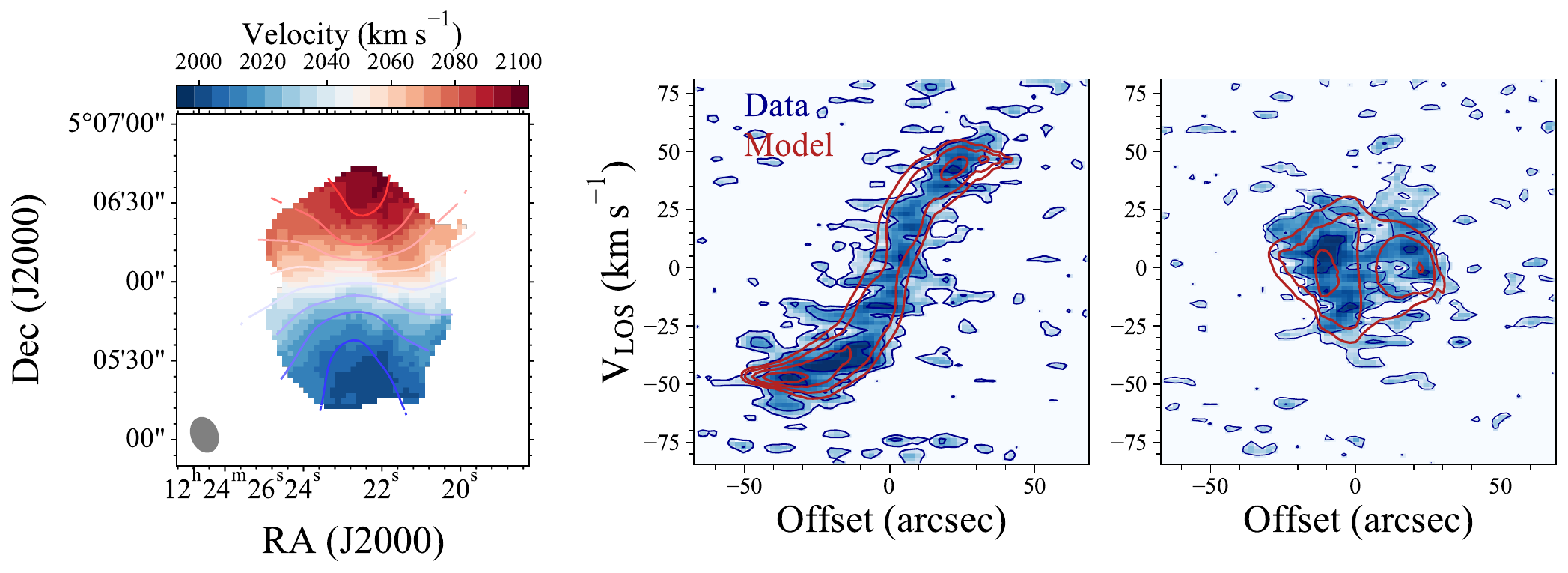}
	\caption{Comparison between the fitted $\rm ^{3D}$Barolo model and the observational data. Left: the observed $\hi$ velocity field is overlaid with model velocity contours from the $\rm ^{3D}$Barolo algorithm. Middle and Right: Kinematical major and minor-axis PV diagram. The offsets are measured from the $\hi$ centre along the major or minor axes, respectively. The observed data are shown in blue with the red contours showing the best-fitting model.}
	\label{fig:VCC693_barolo_fitting.pdf}
\end{figure*}

The best-fit velocity field is overlaid as contours on the $\hi$ {\sc moment1} map in the left panel of Figure \ref{fig:VCC693_barolo_fitting.pdf}, and the position-velocity (PV) diagrams along the major axis and minor axis are shown respectively in the middle and right panels.
$\rm ^{3D}$Barolo provides a reasonably good fit to the velocity field of this galaxy. 
The model even captures the Z-shaped twist in the inner region.
The major-axis PV is consistent with that of a typical rotating disk, showing a flat velocity profile in the outer region.
This suggests that, despite the significant disturbances found in the optical images, the $\hi$ disk has settled into a regularly rotating disk.
A twist appears in the isovelocity contours of the outer disk near the minor axis. 
Therefore, we investigate possible radial flow motions, as shown in Appendix \ref{sec:mass flow rates}, but find no significant gas flow within uncertainties.

\subsection{$\hi$ rotation curves}

\begin{figure}[htbp]
	\centering
	\includegraphics[width=1\linewidth]{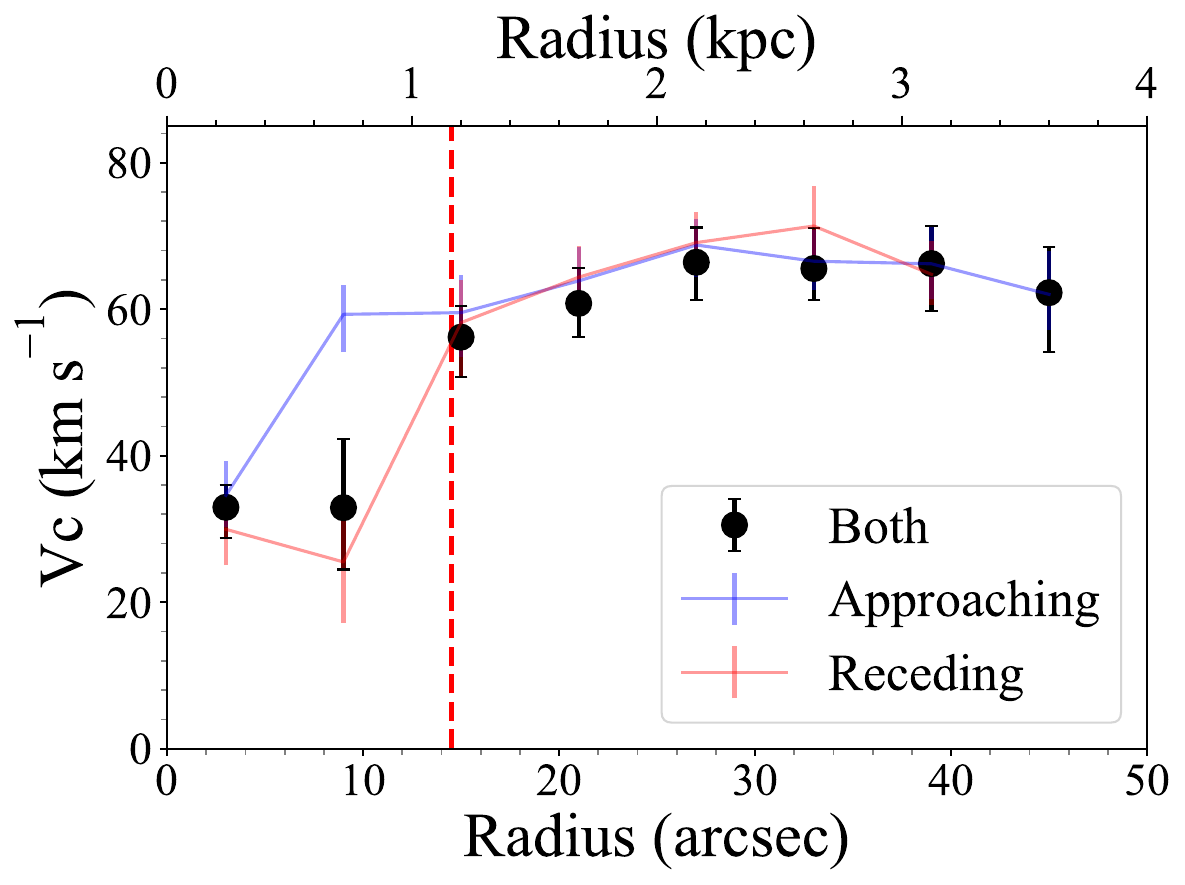}
	\caption{$\hi$ circular velocity profiles of VCC 693 derived from $\rm ^{3D}$Barolo for the receding side (red), approaching side (blue) and the full disk (black dots). The red dashed vertical line indicates the radius of the stellar bar.}
	\label{fig:rotation_curves.pdf}
\end{figure}

Figure \ref{fig:rotation_curves.pdf} shows the best-fit $\hi$ rotation curves.
The rotation velocities of the approaching and receding sides agree with each other across most of the disk, except in the radial range of $\sim$ 0.5-1.5 kpc, where the off-centre stellar bar induces a Z-shaped twist in the velocity field.
The $\hi$ rotation curve is derived out to a radius of $\sim45''$ (3.7 kpc) and the rotation velocity in the flat part of the curve is derived by taking the median values across the three outer rings ($V_{\rm circ}$ = 67.0$^{+2.9}_{-2.8}$ $\rm km\,s^{-1}$, excluding the slightly declining outermost ring).
We estimated the dynamical mass ($M_{\rm dyn}=3.8^{+0.3}_{-0.4}\times 10^{9}\,\rm M_{\odot}$) within the outermost ring using $M_{\rm{dyn}} = V_{\rm{circ}}^{2}r_{\hi}/G$.
Additionally, we compared VCC 693 with other galaxies on the baryonic Tully-Fisher relation.
The fitted velocity (67.0$^{+2.9}_{-2.8}$ $\rm km\,s^{-1}$) is larger than the predicted velocity (55.0$\pm 3$ $\rm km\,s^{-1}$) from \cite{McNichols2016}, but is consistent with the relation when measurement uncertainties, particularly in inclination, are considered. 

\subsection{The dark matter halo density profile}\label{subsec:dark matter}
\begin{figure*}[htbp]
	\centering
	\includegraphics[width=1\linewidth]{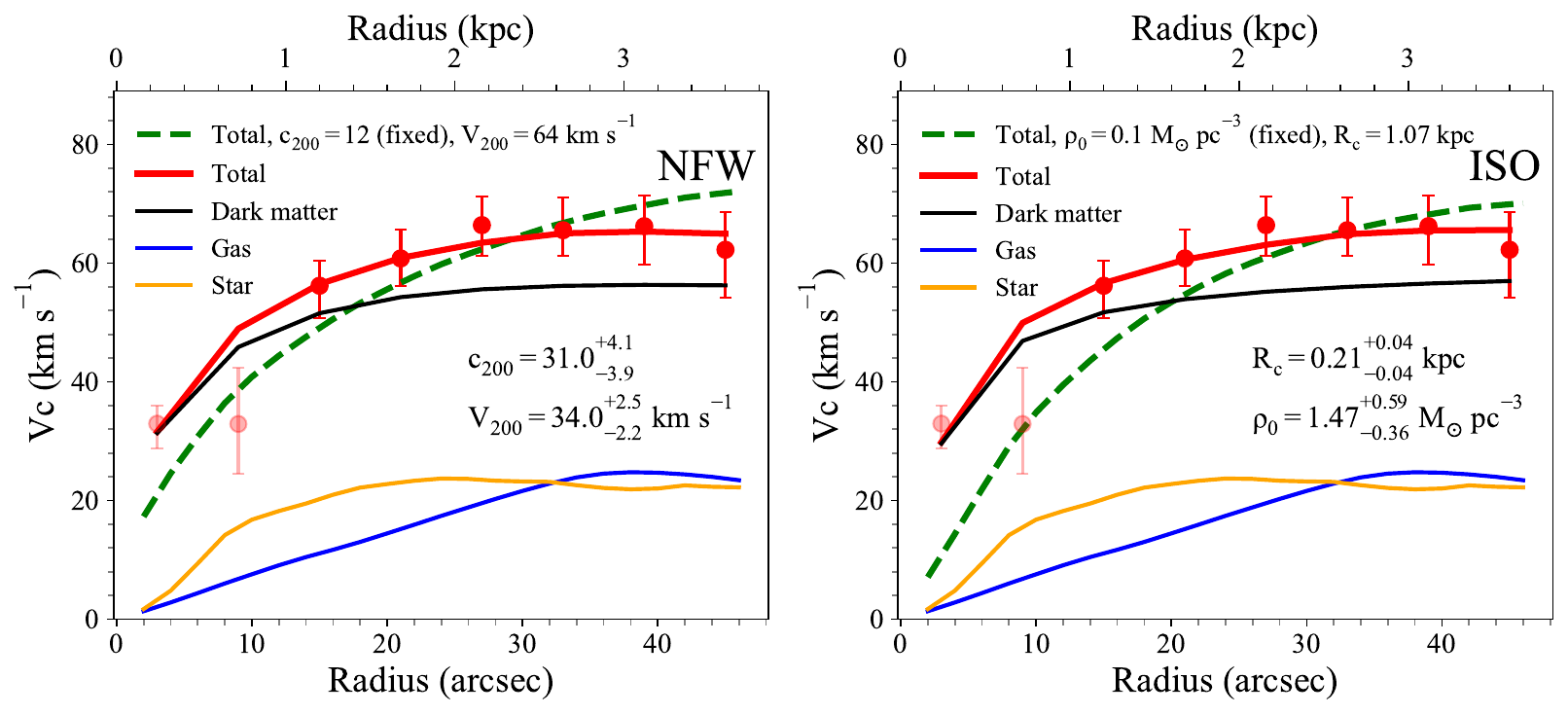}
	\caption{Decomposition of the observed rotation curve of VCC 693 using the NFW (left panel) and spherical pseudo-isothermal (ISO, right panel) dark matter halo models. The red points and red lines represent the observed circular velocity and best-fitting rotation curve. The first two light red points indicate data affected by the bar, but including them in the fitting does not significantly affect the fitted parameters. The blue, orange, and black lines show the contribution of the stellar, gas, and dark matter components to the total circular velocity, respectively. The best-fitting rotation curves with $c_{200}$ fixed at 12 in the left panel and $\rho_{0}$ fixed at $0.1\,\rm M_{\odot}\,pc^{-3}$ in the right panel, are over-plotted with green dashed lines.}
	\label{fig:NFW_fitting.pdf}
\end{figure*}

With the surface mass profiles of baryonic components in hand (see Figure \ref{fig:VCC693_density_profile.pdf}), a disk-halo decomposition can be performed to derive the dark matter distribution of VCC 693.
The observed circular velocity can be dynamically decomposed into three mass components:
\begin{eqnarray}
V_{\rm{c}}^{2} = V_{\rm{\star}}^{2}+V_{\rm{gas}}^{2}+V_{\rm{DM}}^{2},
\end{eqnarray}
where $V_{\rm{\star}}$ and $V_{\rm{gas}}$ represent the gravitational contributions of the stars and gas based on their respective mass distributions, $V_{\rm{DM}}$ is the contribution from the dark matter halo.
We followed the method in \citet{Oh2015} to calculate the corresponding rotation velocities of the gas and stellar components. The method has been widely used to estimate the dynamical contributions of stellar and gaseous components and to test dark matter halo models in galaxies \citep{deBlok2008,Oh2011,Pina2024}.
Briefly, we fit the inclination-corrected surface profiles of each baryonic component (Figure \ref{fig:VCC693_density_profile.pdf}) with a disk mass model and then computed the corresponding rotation velocity as a function of radius along the disk, based on the resulting gravitational potential. In particular, the stellar disk is modelled assuming a vertical sech$^2$ scale height distribution with a ratio of the radial scale length to vertical scale height (h/z) of 5, which is a reasonable assumption for dwarfs \citep{Kregel2004}. For the gaseous component, we adopted an infinitely thin and axisymmetric gas disk \citep{Bosch2001}.
The resulting gas and stellar rotation velocities are computed with the {\sc rotmod} task in \texttt{GIPSY} \citep{Hulst1992}.
The resulting dynamical masses of the gaseous and stellar components are $\rm \sim 2.8 \times 10^{8}\,M_{\odot}$ and $\rm 3.0 \times 10^{8}\,M_{\odot}$, respectively, in broad agreement with those in Table \ref{tab:property}.

Then we subtracted the equivalent rotation velocities contributed by the gas and stellar components from the total rotation velocities (in quadrature) shown in Figure \ref{fig:rotation_curves.pdf}, and obtained the rotation curves contributed by the dark matter halo.
To parameterize the dark matter halo profile, we considered both the NFW model \citep{Navarro1997} and the spherical pseudo isothermal (ISO) model \citep{Begeman1991}.
The density profile of NFW is given by 
\begin{eqnarray}
\rho_{\rm NFW}(r)=\frac{4\rho_{s}}{\frac{r}{r_{s}}+(1+r/r_{s})^{2}},
\end{eqnarray}
and the dark matter halo rotation velocity is given by,
\begin{eqnarray}
V_{\rm NFW}(r)=V_{200}\sqrt{\frac{\ln(1+c_{200}x)-c_{200}x/(1+c_{200}x)}{x[\ln(1+c_{200})-c_{200}/(1+c_{200})]}}.
\end{eqnarray}
where $r_{\rm s}$ and $\rho_{s}$ are the scale radius and the density at $r_{\rm s}$.
The concentration parameter $c_{200}$ is defined as $r_{200}/r_{\rm s}$, and $V_{200}$ is the rotation velocity, where $r_{200}$ is the virial radius.
The variable $x$ is defined as $r/r_{200}$.

Compared with the NFW model, the ISO model has a constant density core and the density profile is given by:
\begin{eqnarray}
\rho_{\rm ISO}(r)=\frac{\rho_{0}}{1+(r/r_{c})^{2}},
\end{eqnarray}
where $\rho_{0}$ and $r_{c}$ are the core density and radius of the dark matter halo, respectively.
The corresponding rotation velocity is given by,
\begin{eqnarray}
V_{\rm ISO}(r)=\sqrt{4\pi G \rho_{0}r_{c}^{2}[1-\frac{r_{c}}{r} \arctan(\frac{r}{r_{c}})]}.
\end{eqnarray}

We use the Markov chain Monte Carlo (MCMC) technique implemented in \texttt{EMCEE} \citep{Foreman-Mackey2013} to sample the posterior parameter distributions of the dark matter halo parameters, and the uncertainties are taken at the 16th and 84th percentiles.
The posterior distributions of the fitted parameters are shown in Figure \ref{fig:MCMC_all.png} in Appendix \ref{sec:mcmc posterior}.
Results of the best-fit mass decomposition are shown in Figure \ref{fig:NFW_fitting.pdf}.
The left panel shows the results based on the NFW dark matter halo profile, while the right panel shows the results by adopting the ISO dark matter halo profile.
The radius range affected by the off-centre bar has been excluded from the fitting. 
The quality of fitting based on the two dark matter halo profiles is similarly good, suggesting that our observations cannot distinguish between the two halo models. 
For the NFW model, the derived concentration parameter $c_{200}$ is 31.0$^{+4.1}_{-3.9}$, which is unusually high for a dwarf galaxy of this mass ($c_{200}\sim10$) \citep[e.g.,][]{Oh2015,Pina2025}.
Such a high $c_{200}$ value suggests a very concentrated dark matter halo of VCC 693.
Uncertainty in the baryonic mass (e.g., mass-to-light ratio and molecular mass) could increase the halo concentration, but we expect it to play only a secondary role in dwarf galaxies.
Although it remains unclear whether mergers increase the halo concentration, a cuspy halo tends to retain its cuspy profile after mergers, as long as at least one of the progenitor galaxies had a cuspy halo \citep{Boylan-Kolchin2004}.
Furthermore, \citet{WangK2020} had found that the concentration parameters could be dramatically changed during the merger, which could explain the high $c_{200}$ observed in VCC 693.
Meanwhile, the results obtained from the ISO model fitting lead to similar results, with the density (1.47$^{+0.59}_{-0.36}$ $\rm M_{\odot}\,pc^{-3}$) and radius (0.21$^{+0.04}_{-0.04}$ kpc) of the core being much higher than those of ordinary dwarfs with similar maximum rotation velocities.
This is in a similar vein of the unusually high concentration for the NFW model. 
To demonstrate the robustness of the fitting results of the unusually high concentration (for the NFW model) or core density (for the ISO model), we ran the mass profile modelling by fixing $c_{200} = 12$, which is the typical value for normal galaxies of comparable mass for the NFW halo profile, and by fixing $\rho_{0}=0.1\,\rm M_{\odot}\,pc^{-3}$ for the ISO halo profile. 
The best-fitting rotation curves from these two test runs are over-plotted with green dashed lines in the left and right panels of Figure \ref{fig:NFW_fitting.pdf}, respectively. 
It is evident that the two constrained fittings fail to reproduce the observed steeply rising and subsequently flat rotation curve over the observed radial range.
Furthermore, through systematic testing, we found that excluding the first two points of the rotation curve (especially the second point, which has a large uncertainty) from the dark matter fitting does not significantly affect the fitting results.
Lastly, we note that the above results would not change noticeably if a larger distance of VCC 693 (e.g. the fiducial distance predicted by TFR) is adopted. 

\section{Reproducing the major features of VCC 693 via a damp dwarf galaxy merger}\label{sec: Discussion}

The analysis presented above suggests that the stellar distribution of VCC 693 is dominated by complex tidal features spanning the entire system. Such substantial disruption could result from either a near-major galaxy merger or multiple high-speed fly-by interactions \citep[i.e., galaxy harassment;][]{Moore1996} with galaxies.

In the fly-by scenario, our simulations (not shown here) indicate that an intruder galaxy with a mass lower than or comparable to that of the target galaxy has only a limited impact on the target's stellar structure. We therefore focus on interactions involving an intruder that is significantly more massive than the target galaxy. It is worth noting that the nearly normal $\hi$ gas richness of VCC 693, together with its location on the periphery of the Virgo cluster, implies that it is at most a recent infaller and is therefore unlikely to have experienced repeated high-speed encounters typical of a cluster core environment \citep{Smith2010,Smith2015}.

In the merger scenario, not only do the stellar components but also the gas components of the progenitor galaxies play a significant role in shaping the properties of the merger remnant.
One or both progenitors of VCC 693 must have been gas-rich.
Early simulations have shown that mergers between gas-bearing, star-dominated spiral galaxies typically result in significant gravitational torque-driven gas inflow and central starbursts \citep{Barnes1996, Mihos1996}. Furthermore, later simulations \citep[e.g.,][]{Springel2005} suggest that the magnitude of gas inflow and the resulting central starburst may be suppressed in high-gas-fraction mergers, as is likely the case for VCC 693. Nevertheless, hydrodynamical simulations of gas-rich dwarf galaxy mergers remain rare in the literature \citep{Zhang2020, SunW2025}. 
In the following, we attempt to use such simulations to shed light on the progenitor mass ratio of VCC 693 and the gas richness of its two progenitors, based on a qualitative comparison with the observed properties of VCC 693. 
We emphasize, however, that these simulations are not intended to reproduce the detailed spatial distributions of the galaxy's fine structures, as doing so would require an extensive exploration of collision parameters beyond the scope of this work.

\subsection{Setup of the simulations}\label{subsec:Simulation}
\begin{figure*}[htbp]
	\centering
	\includegraphics[width=1\linewidth]{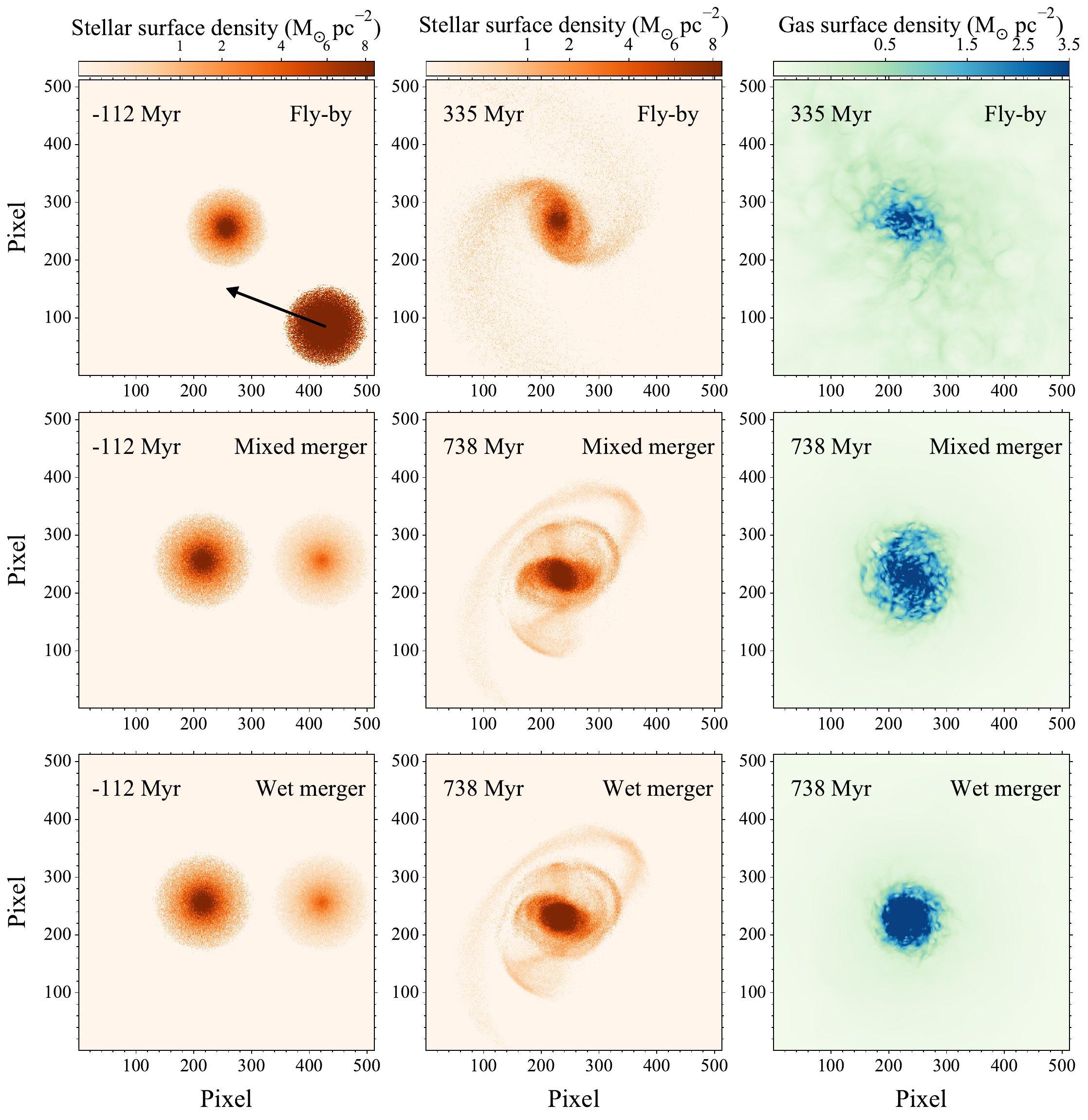}
	\caption{Projected surface density distribution of stellar (orange) and gas (blue) particles. Top: the fly-by scenario; Middle: the mixed merger scenario; Bottom: the wet merger scenario. The left panels show the stellar snapshots at the beginning of the simulation. The middle and right panels display selected snapshots highlighting the special structures. The simulations are performed with a spatial resolution of $\sim$50 pc. We define the first pericentric passage of the two galaxies in the merger scenario as the zero point in time. In the fly-by scenario, the black arrow marks the trajectory of the massive galaxy.}
	\label{fig:VCC693_simulation.pdf}
\end{figure*}

\begin{figure}[htbp]
	\centering
	\includegraphics[width=1\linewidth]{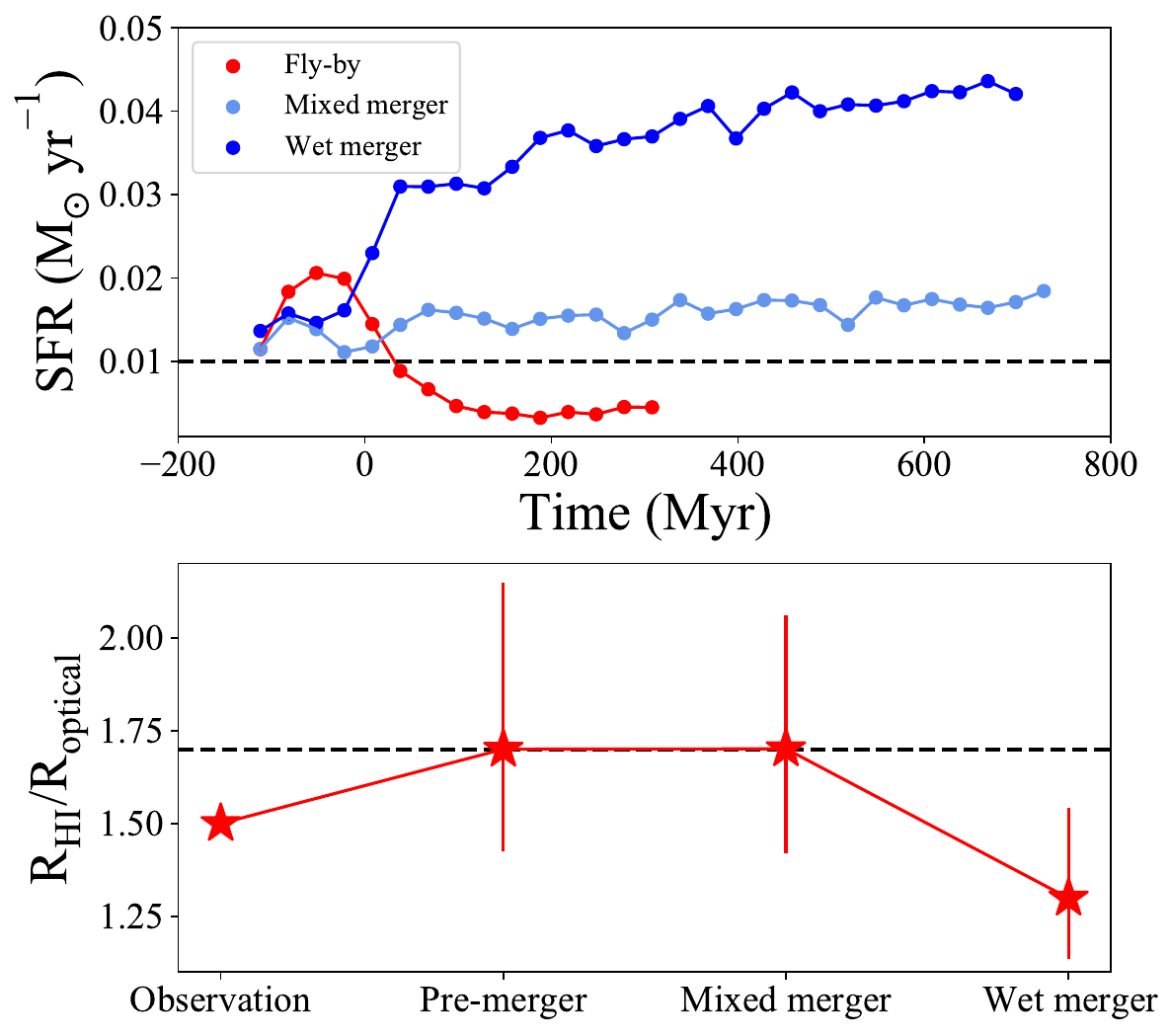}
	\caption{Comparison of the SFR (top) and $\rm R_{\hi}/R_{optical}$ (bottom) among the three scenarios. The top panel shows the evolution of the SFR over time for the fly-by (red), mixed merger (light blue), and wet merger (blue) simulations. We define the first pericentric passage of the two galaxies in the merger scenario as the zero point in time. The black dashed lines represent the values from the isolated disk simulation. The bottom panel shows the measured $\rm R_{\hi}/R_{optical}$ values for observations of VCC 693 (left) and for the simulations (right three points: pre-merger, mixed merger, and wet merger). The optical radius of observations and typical value of late-type galaxies (black dashed line) are measured from the 25 mag arcsec$^{-2}$ isophote of the optical $B$-band image, while in the simulations, we adopt the optical radius corresponding to a pre-merger $\rm R_{\hi}/R_{optical}$ of 1.7 as the reference ($\rm \log \Sigma_{star}=0.48\, M_{\odot}\,pc^{-2}$), with error bars indicating a $\pm 0.2$ dex range in $\rm \log \Sigma_{star}$. }
	\label{fig:VCC693_simulation_SFR.pdf}
\end{figure}

We explore three scenarios through simulations: a single fly-by interaction with a massive intruder, a mixed galaxy merger between a gas-rich primary and a gas-free secondary dwarf galaxy (hereafter mixed merger), and a wet merger between two gas-rich dwarfs. 
The simulations are performed using the adaptive mesh refinement code \texttt{RAMSES} \citep{Teyssier2002}. 
The initial setup of the galaxy models---comprising exponential disks embedded in NFW dark matter halos---as well as the implementation of gas cooling, star formation, and feedback, follows the methodology described in \citet{SunW2025}. The simulations employ a box size of 50 kpc and a spatial resolution of 50 pc, which is sufficient to contain the model galaxy disks and to resolve their structural variations.

For the fly-by simulation, we set a massive galaxy (the intruder) at an initial offset of 10 kpc in the x and y directions, performing a fly-by at a velocity of 500 km s$^{-1}$ with respect to the target dwarf galaxy.
The target dwarf galaxy has a stellar mass of M$_\star$=1$\times10^{8}\,\rm M_\odot$ and a dark matter halo of M$_{200}$=1$\times10^{11}\,\rm M_\odot$ and a halo concentration of 7, which is in agreement with halo abundance matching \citep{Behroozi2014}.
The disk is set to be thick, with an exponential scale-length of 0.5 kpc and a vertical to radial scale ratio of 0.6 \citep{Sanchez-Janssen2010}.
The massive galaxy is structurally similar to the target dwarf but 10 times more massive.

For the mixed merger simulation, our fiducial model has a 4:1 mass ratio between a primary dwarf with 1$\times10^{8}\,\rm M_\odot$ gas mass and a gas-free secondary dwarf.
We use the model galaxy for the primary dwarf with M$_\star$=1$\times10^{8}\,\rm M_\odot$, whereas the secondary dwarf has M$_\star$=2.5$\times10^{7}\,\rm M_\odot$.
In the wet merger simulations, with the same stellar setup as the mixed merger, a gas disk is added to the secondary dwarf, with a gas-to-stellar mass ratio equal to that of the primary dwarf (i.e., 1:1). In the merger simulations, we adopt prograde interactions occurring in the plane of the primary galaxy’s disk. This configuration can naturally generate coherent tidal features within the disk plane through resonant coupling with the disk motion, and ensures that tidal debris stripped from the progenitors can wrap around the disk when viewed face-on, like what is observed in VCC 693.

\subsection{Comparison of simulation results with VCC 693}
We display the projected surface density distributions of stellar and gas particles at selected snapshots for the fly-by and merger simulations in Figure \ref{fig:VCC693_simulation.pdf}, and the evolution of the SFR over time and $\hi$-to-stellar radius ratio for the three scenarios in Figure \ref{fig:VCC693_simulation_SFR.pdf}. 
In the following, we briefly discuss the fly-by and the two merger simulations and their relevance to VCC 693.

\subsubsection{Fly-by interaction with a massive galaxy}
Following the high-speed fly-by interaction with a massive galaxy, the target dwarf galaxy develops a disturbed stellar distribution, with tidal features dominated by two symmetric tidal arms. 
The overall stellar distribution is axially symmetric.
This outcome is similar to that seen in previous simulations for massive galaxies \citep{Moore1996}.
However, the resulting stellar morphology differs from that of VCC 693, which displays a far more complex array of tidal features without dominant tidal arms.
Moreover, the fly-by interaction produces a very extended gas distribution, shown in the top-right panel of Figure \ref{fig:VCC693_simulation.pdf}, again inconsistent with the $\hi$ distribution of VCC 693, which is only slightly more extended than the stellar component. 
Lastly, the SFR is briefly enhanced by a factor of 2 during the interaction but then quickly falls well below that of a simulated dwarf galaxy without experiencing a fly-by interaction (blue dashed line in the top panel of Figure \ref{fig:VCC693_simulation_SFR.pdf}), which is again inconsistent with the largely normal SFR of VCC 693 for its mass.

We note that the two massive galaxies projected close to VCC 693---VCC 613 (projected separation: 71 kpc; radial velocity difference: $-$381 km s$^{-1}$) and VCC 785 (projected separation: 116 kpc; radial velocity difference: 506 km s$^{-1}$)---show no obvious signatures of tidal interactions in either their stellar or gaseous distributions (Appendix \ref{sec:neighbor_HIspectra}). 
Our FAST $\hi$ mapping also reveals no tidal streams between them and VCC 693. 
Therefore, the apparent proximity of the three galaxies most likely results from line-of-sight projection rather than physical association.

\subsubsection{Merger scenarios}

According to \citet{Conselice2022}, the merger timescale at z $\sim0$ ranges from $\sim\,1.25$ to 2 Gyr for objects with stellar masses of $\rm M_{\star}>10^{9}\,M_{\odot}$.
For mergers between dwarf galaxies, the merger timescale may be shorter. 
We find that the simulated stellar surface density distribution between $\sim$ 740 and 1200 Myr after the first pericentric passage exhibits tidal features similar to those observed in VCC 693 (see the middle and bottom panels of Figure \ref{fig:VCC693_simulation.pdf}).
In particular, the merger simulation produces broad tidal arms emanating from the primary galaxy, as well as several faint tidal arms closely wrapping around the outskirts.
Meanwhile, a bar-like structure is also found at the centre of the galaxy, although it appears more symmetric compared to the bar in VCC 693.
The stellar components in the two merger simulations do not show significant differences. 
We have performed two additional mixed-merger simulations (not shown here) with mass ratios of 2:1 and 5:1, respectively. 
We find that in the 2:1 case the target galaxy is fully disrupted soon after the first pericentric passage, leaving no prominent tidal arms emanating from the primary, while a 5:1 mass-ratio merger results in a regularly shaped stellar distribution, which is again inconsistent with that observed in VCC 693. 
While we do not attempt to pin down the exact primary-to-secondary progenitor mass ratio for VCC 693 through simulations, our favored ratio lies between 3:1 and 4:1.

By $\sim$ 700 Myr after the first pericentric passage in the two 4:1 merger simulations, the gas components have settled into a regularly rotating disk with no noticeable signatures of tidal interaction, similar to that observed in VCC 693. 
The gas in the mixed-merger simulation settles a couple of hundred million years earlier than in the wet-merger case. 
In addition, the gas in the wet-merger simulation is more centrally concentrated than that in the mixed-merger simulation, as shown in the right panels of Figure \ref{fig:VCC693_simulation.pdf}.

The top panel of Figure \ref{fig:VCC693_simulation_SFR.pdf} compares the star formation histories of the two merger simulations and the fly-by simulation. 
The two merger simulations exhibit a persistent and moderate enhancement of SFR following the first pericentric passage, with the effect being more pronounced in the wet merger. 
In particular, the wet merger simulation reaches a SFR approximately four times higher than that of the primary progenitor prior to the merger, consistent with the well-established results for wet mergers in the literature \citep[e.g.,][]{Mihos1994,Luo2014}. 
The global SFR of VCC 693 is consistent with the typical values of ordinary galaxies of similar stellar mass, lying within the 1$\sigma$ scatter of the main-sequence relation \citep{Dale2023}. 
Taken at face value, the very mild enhancement of the SFR in the mixed-merger simulation appears to be in better agreement with the observations.
Nevertheless, we emphasize that, on the one hand, modelling star formation remains notoriously challenging even in state-of-the-art simulations; on the other hand, star formation during galaxy mergers is expected to vary on timescales much shorter than those of stellar tidal features, which complicates a direct comparison with the current SFR of VCC 693.

The bottom panel of Figure \ref{fig:VCC693_simulation_SFR.pdf} shows the measured $\rm R_{\hi}/R_{optical}$ ratios for VCC 693 (left star symbol) and for the simulations (right three stars). 
For VCC 693, the $\hi$ radius is measured at the N($\hi$) = 1 $\rm M_{\odot}\,pc^{-2}$ contour level, while the optical radius is defined as the $B$-band 25 mag arcsec$^{-2}$ isophote, following the standard definitions in the literature. 
In the simulations, the $\hi$ radius is defined in the same way as in the observations. 
The optical radius is measured at a stellar mass surface density of $\rm \log\Sigma_{star}=0.48\, M_{\odot}\,pc^{-2}$, which corresponds to $\rm R_{\hi}/R_{optical}=1.7$ (the typical value for nearby late-type galaxies) in the pre-merger configuration. 
We consider a $\pm 0.2$ dex uncertainty in the fiducial $\log \Sigma_{\rm star}$ to estimate the uncertainties of the radius ratios.
The $\rm R_{\hi}/R_{\rm optical}$ of VCC 693 is only 0.2 dex lower than the typical value. 
In the simulations, the mixed merger has nearly the same radius ratio to the pre-merger primary galaxy, whereas the wet merger produces a slightly larger optical radius and a significantly smaller $\hi$ radius, resulting in an $\rm R_{\hi}/R_{\rm optical}$ of 1.30 that is 0.40 lower than in the pre-merger progenitor. 
Taken at face value, the observed ratio of VCC 693 lies between the predicted values of mixed and wet merger simulations.

Based on the above comparisons, the favoured scenario for VCC 693 is a near-major merger between a gas-rich primary and a relatively gas-poor secondary (which we define as a damp merger). 
However, we acknowledge that it is difficult to draw a definitive conclusion regarding the gas richness of the secondary progenitor, given the limitations of the existing observations and the restricted exploration of parameter space in the simulations.

\section{Summary and discussion}\label{sec: Summary}
We have performed a comprehensive study of the near-major dwarf merger remnant, VCC 693, located in the outskirts of the Virgo cluster. 
Our analysis is based on multi-wavelength data, including optical imaging from the NGVS and VESTIGE surveys, fiber spectroscopy from SDSS and IFU spectroscopy from MUSE, and $\hi$ 21 cm mapping from the VLA and FAST as part of the AVID project. 
This represents one of the first resolved case studies of near-major post-merger of late-type dwarf galaxies, following VCC 848 \citep{Zhang2020a,Zhang2020} and VCC 479 \citep{SunW2025}.

VCC 693 is distinguished by its complex optical tidal structures spanning across the whole system, including a series of faint tidal arms wrapping around the main body, incomplete ring-like tidal arms hosting young stellar populations, an off-centre stellar bar, and a broad tidal arm emanating from the centre (Section \ref{sec: stellar properties}). 
In contrast, its $\hi$ gas has largely settled into a regular rotating disk with a normal $\hi$ surface density profile. 
Based on a qualitative comparison with hydrodynamical simulations, the overall observational properties of VCC 693 can be explained by a coalesced near-major merger between two late-type dwarf galaxies with a primary-to-secondary stellar mass ratio of $\sim$ 3-4. 
While we cannot draw a definitive conclusion, our analysis favours a damp merger scenario involving a gas-rich primary and a relatively gas-poor secondary. 

The integrated SFR, central brightness, and surface brightness profiles are all in good agreement with those expected for normal dwarf galaxies of similar stellar mass, suggesting a very limited degree of structural transformation induced by the merger event. The central gas-phase metallicity is slightly elevated by approximately 0.2 dex than the expected typical value, which is in line with a mild enhancement of star formation efficiency during the merger event. Nevertheless, a decomposition of the $\hi$ rotation curves into baryonic and dark matter components suggests that the dark matter halo has an unusually high concentration ($c_{200}\approx 31$) for an NFW profile, or an unusually high core density ($\sim$1.47 $\rm M_{\odot}\,pc^{-3}$) for an ISO profile (Section \ref{subsec:dark matter}), which may be attributed to the violent relaxation in a merger between galaxies with comparable dark matter halo masses. 

VCC 693 has an $\hi$ mass and an $\hi$-to-optical size ratio that are slightly lower than---but still within the 1$\sigma$ scatter of---the typical values for normal galaxies of comparable mass, indicating that VCC 693 is at most a recent infaller into the outskirts of the Virgo cluster and has not experienced significant environmental stripping. 

With a relatively high galaxy number density and a moderately high velocity dispersion, infalling galaxy groups and the outskirts of galaxy clusters are favourable environments for galaxy mergers. In such regimes, galaxies may experience a wide range of environmental effects, and damp/mixed mergers are most likely to occur here, constituting an integral part of the so-called galaxy pre-processing \citep{Fujita2004}. Indeed, all three late-type near-major dwarf-dwarf merger remnants studied so far---including VCC 848, VCC 479, and VCC 693---appear to involve at least one progenitor with a relatively low gas fraction, in line with the speculation above. Moreover, all three merger remnants exhibit stellar structural properties that are indistinguishable from those of ordinary dwarf galaxies of similar mass, suggesting that such mergers play only a limited role in the structural transformation of dwarf galaxies.

\begin{acknowledgements}
This work is supported by the National Science Foundation of China (NSFC, Grant No. 12233008, 12122303, 11973039, 12173079), the National Key R\&D Program of China (2023YFA1608100), the Strategic Priority Research Program of the Chinese Academy of Sciences (Grant No. XDB0550200), the Cyrus Chun Ying Tang Foundations, and the 111 Project for "Observational and Theoretical Research on Dark Matter and Dark Energy" (B23042).
This work is also supported by the China Manned Space Project (grant Nos. CMS- CSST-2021B02 and CMS-CSST-2021-A07, CMS-CSST-2021-A06). We also acknowledge support from the CAS Pioneer Hundred Talents Program, the Strategic Priority Research Program of Chinese Academy of Sciences (grant No. XDB 41000000), and the Cyrus Chun Ying Tang Foundations. This work made use of the data from FAST (Five-hundred-meter Aperture Spherical radio Telescope)(https://cstr.cn/31116.02.FAST). FAST is a Chinese national megascience facility, operated by National Astronomical Observatories, Chinese Academy of Sciences. We are grateful to Professor Jing Wang for her guidance in the design of the FAST observation strategy, and to her student Mr. Yang Dong for helpful suggestions during the observation process.
SHOH acknowledges a support from the National Research Foundation of Korea (NRF) grant funded by the Korea government (Ministry of Science and ICT: MSIT) (RS2022-00197685, RS-2023-00243222).
RS acknowledges financial support from FONDECYT Regular 2023 project No. 1230441 and also gratefully acknowledges financial support from ANID - MILENIO NCN2024\_112
\end{acknowledgements}

\appendix

\renewcommand{\appendixname}{Appendix}
\section{The $\hi$ spectra of neighboring galaxies of VCC 693}\label{sec:neighbor_HIspectra}

\begin{table}[b]
    \centering
    \tabcolsep=8pt
    \renewcommand{\arraystretch}{1.5}
    \caption{$\hi$ properties of neighboring galaxies of VCC 693.\label{tab:Neighboring HI property}}
    \begin{tabular}{c| c c c c| c c c c}
    \hline
        & \multicolumn{4}{c|}{FAST} & \multicolumn{4}{c}{ALFALFA} \\
    Galaxy & $\rm V_{sys}$& $W_{50}$ & $\rm S_{int}$ & $\rm \log M_{\hi}$& $\rm V_{sys}$& $W_{50}$ & $\rm S_{int}$ & $\rm \log M_{\hi}$ \\
     & ($\rm km\,s^{-1}$)& ($\rm km\,s^{-1}$) & (Jy $\rm km\,s^{-1}$) & ($M_{\odot}$)& ($\rm km\,s^{-1}$)& ($\rm km\,s^{-1}$) & (Jy $\rm km\,s^{-1}$) & ($M_{\odot}$) \\
     \hline
VCC 785& 2555.9$^{+0.4}_{-0.4}$ & 338.3$^{+0.7}_{-0.9}$ & 13.1$^{+0.05}_{-0.05}$ & 8.92$^{+0.06}_{-0.06}$ &2555.7$^{+0.4}_{-0.4}$ & 343.6$^{+0.8}_{-0.7}$ & 11.8$^{+0.05}_{-0.05}$ & 8.88$^{+0.06}_{-0.06}$\\
VCC 613& 1668.5$^{+0.2}_{-0.2}$ & 304.1$^{+0.5}_{-0.4}$ & 12.04$^{+0.03}_{-0.03}$ & 8.89$^{+0.06}_{-0.06}$ &1667.4$^{+0.4}_{-0.4}$ & 315.2$^{+0.8}_{-0.8}$ & 9.11$^{+0.05}_{-0.06}$ & 8.77$^{+0.06}_{-0.06}$\\
VCC 764& 1969.5$^{+3.6}_{-3.6}$ & 196.8$^{+6.9}_{-7.8}$ & 0.85$^{+0.02}_{-0.02}$ & 7.74$^{+0.06}_{-0.06}$ &2020.6$^{+2.2}_{-1.5}$ & 150.6$^{+3.2}_{-3.4}$ & 0.98$^{+0.04}_{-0.04}$ & 7.80$^{+0.6}_{-0.6}$\\
    \hline
    \end{tabular}
\end{table}

\begin{figure}[htbp]
	\centering
	\includegraphics[width=1\linewidth]{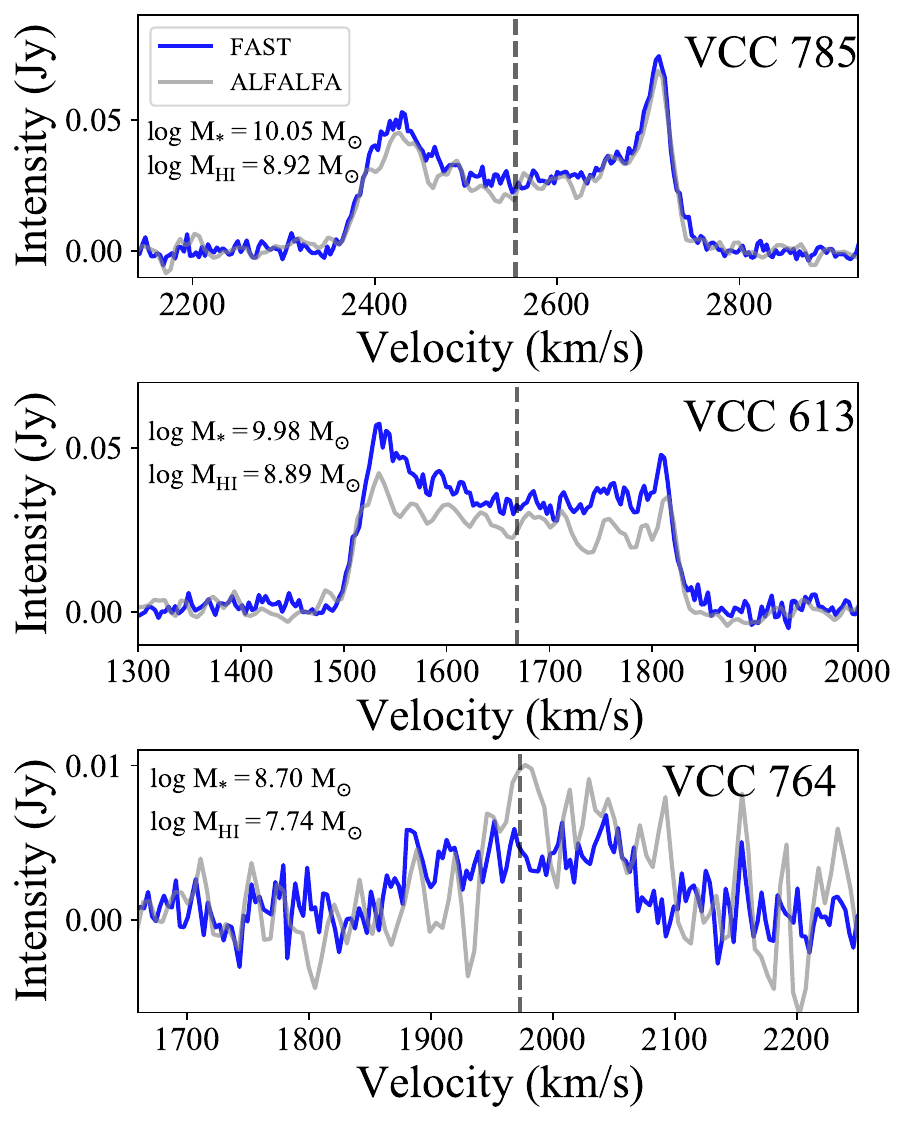}
	\caption{The $\hi$ profiles of neighboring galaxies of VCC 693 from FAST observations, overlaid on the ALFALFA.}
	\label{fig:VCC693_environment_HIspectra.pdf}
\end{figure}

We detected $\hi$ emission from the neighboring galaxies to VCC 693 with FAST. 
The spectra are shown in Figure \ref{fig:VCC693_environment_HIspectra.pdf}. The basic $\hi$ properties are shown in Table \ref{tab:Neighboring HI property}. 
VCC 642 shows no detected $\hi$ emission, and the velocity in Figure \ref{fig:VCC693_environment_1.pdf} comes from the optical spectrum \citep{Zaritsky2023}.
We find a larger $\hi$ mass than the archive ALFALFA data, which we ascribe to  the superior sensitivity of FAST, which enables it to reach a lower detection limit, providing more accurate results than ALFALFA.
They are currently far from the centre of the Virgo cluster and do not exhibit a significant deficiency of their $\hi$ gas, indicating that this is a relatively gas-rich local environment.
We did not find any obvious signatures of environmental effects in the $\hi$ profiles and optical structures of VCC 785 and VCC 613, implying that they may be only recently falling into the outskirts of the Virgo cluster.

\section{The $\hi$ channel map of VCC 693}\label{sec:channel map}
\begin{figure*}[htbp]
	\centering
	\includegraphics[width=0.7\linewidth]{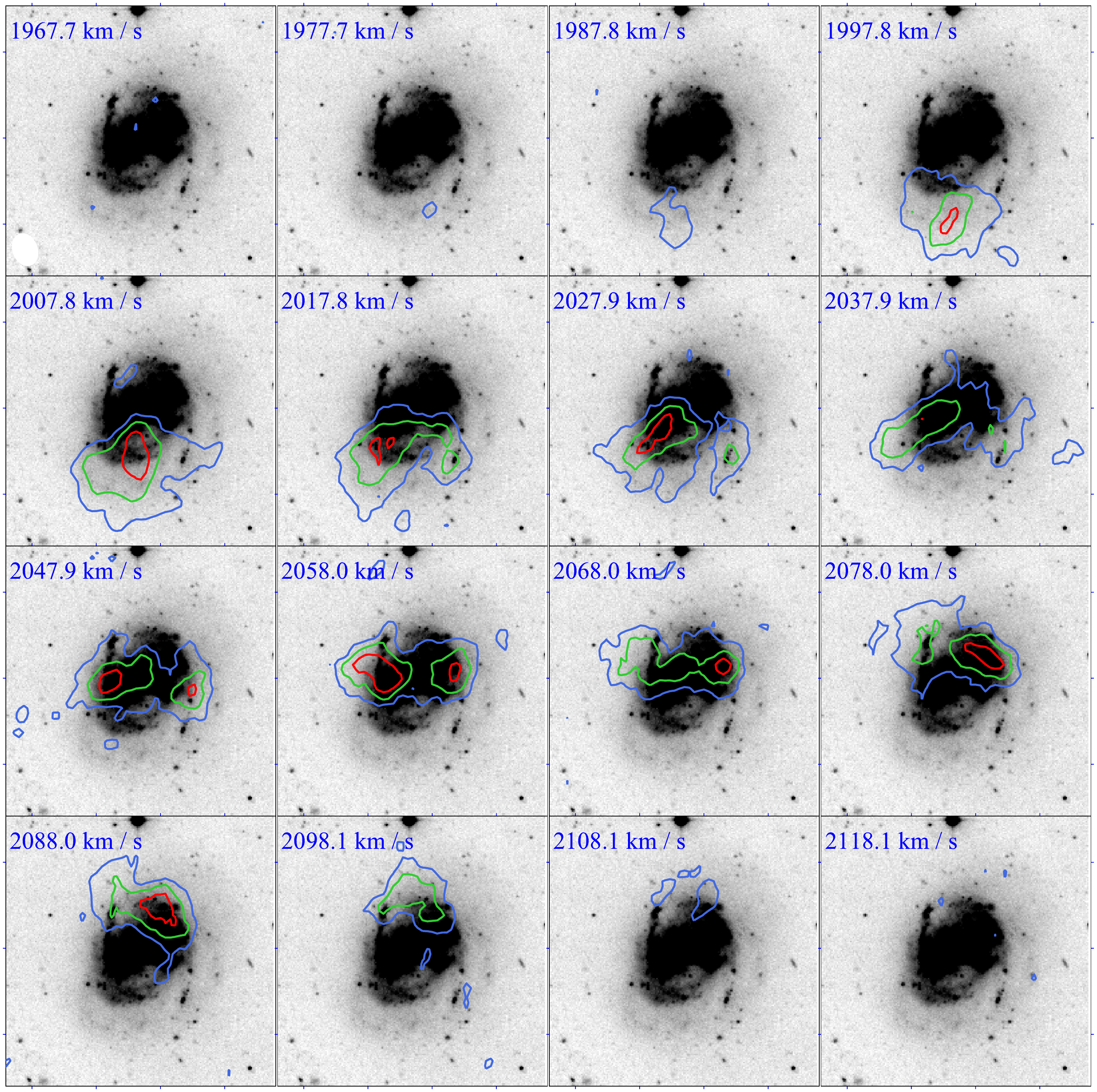}
	\caption{Channel maps based on a natural-weighted VLA data cube of VCC 693. The 3.4 $\rm km\,s^{-1}$ velocity channels were binned here to 10 $\rm km\,s^{-1}$. The blue, green, and red contour indicate 2, 4, and 6$\sigma$, where $\sigma=$ 5.2 mJy beam$^{-1}$ is the rms noise of the data cube. The beam size is indicated in the bottom left of the first panel. The heliocentric velocity is indicated to the top left in each panel. }
	\label{fig:VCC693_channel_map.pdf}
\end{figure*}

Figure \ref{fig:VCC693_channel_map.pdf} shows individual channel maps of the $\hi$ emission from 1967.7 $\rm km\,s^{-1}$ to 2118.1 $\rm km\,s^{-1}$.
The channel maps are based on the natural-weighted $\hi$ data cubes, re-gridded to a channel width of 10 $\rm km\,s^{-1}$.

\section{Radial flows and mass flow rates}\label{sec:mass flow rates}
\begin{figure*}[htbp]
	\centering
	\includegraphics[width=0.9\linewidth]{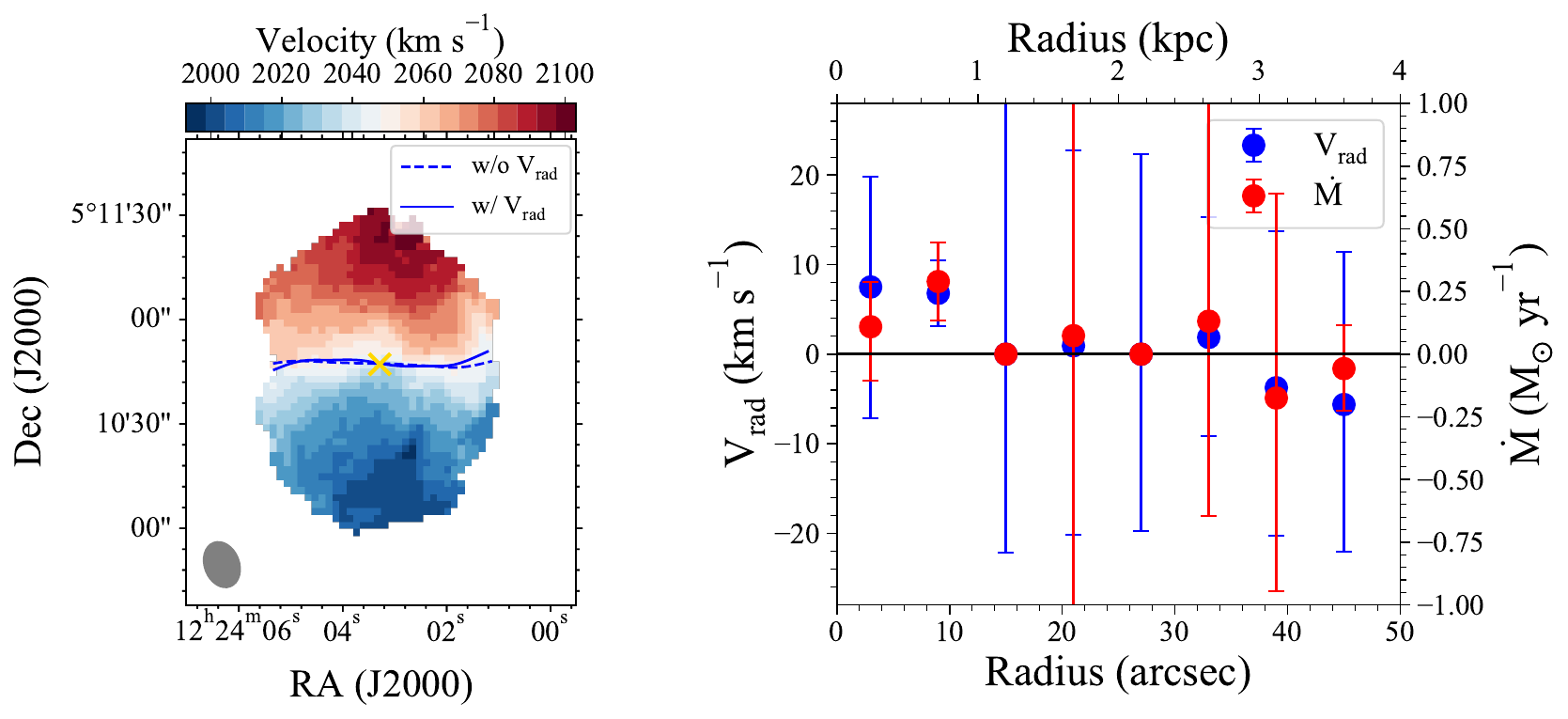}
	\caption{Left panel: the comparison of systemic velocities between cases with (solid line) and without (dashed line) radial velocity fitting. Right panel: the radial velocity (blue) and mass flow rate (red) determined by the $\rm ^{3D}$Barolo fitting on the VLA natural-weighted observations.}
	\label{fig:VCC693_Vrad.pdf}
\end{figure*}

Unlike disk warps which may cause twists of isovelocity contours near both the major and minor axes, radial flows generally introduce twists of isovelocity contours only near the minor axis \citep{DiTeodoro2021}. 
In the $\rm ^{3D}$Barolo fitting process, we carefully modelled the radial motions from the twist of the kinematical minor axis, with the final results presented in Figure \ref{fig:VCC693_Vrad.pdf}.
The $\hi$ mass flow rates are calculated from the $\hi$ surface-density profiles $\rm \Sigma_{\hi}$ and $V_{\rm rad}$ \citep{DiTeodoro2021}: 
\begin{eqnarray}
\dot{M}_{\hi}(R) = 2\pi \Sigma_{\hi}(R)V_{\rm rad}(R).
\end{eqnarray}
In the left panel, we demonstrate the impact of radial flow on the modelling of the velocity field, where the solid and dashed lines represent the systemic velocity with and without radial motion fitting, respectively.
The model with radial motions matches the observed velocity field better than that without radial motion, although the difference is not significant. 
The radial motions and mass flow rate extracted from the $\rm ^{3D}$Barolo are shown in the right panel.
As can be seen, no significant gas flow is detected within uncertainties.

\section{Posterior distribution of the dark matter model}\label{sec:mcmc posterior}
\begin{figure*}[htbp]
	\centering
	\includegraphics[width=1\linewidth]{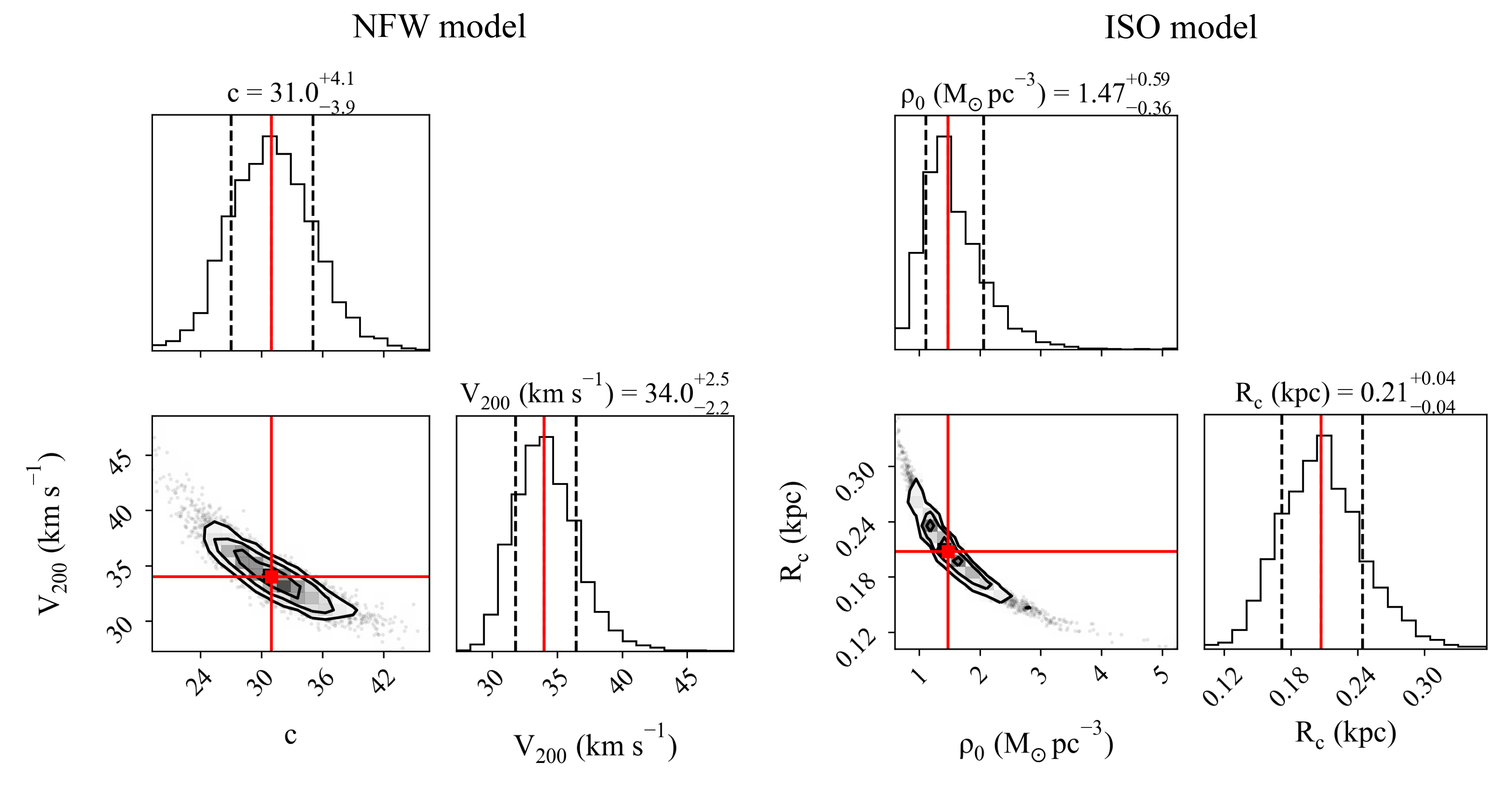}
	\caption{Posterior distributions of parameters of the NFW (left) and ISO (right) models fitted to VCC 693. The red squares indicate the maximum-likelihood values. The black contours represent the 16th, 50th, and 84th percentiles of the distribution. In each histogram, the red vertical line indicates the maximum-likelihood value, while the two black dashed lines mark the 16th and 84th percentile confidence intervals.}
	\label{fig:MCMC_all.png}
\end{figure*}

Figure \ref{fig:MCMC_all.png} presents the posterior distributions of the fitted parameters for the NFW (left) and ISO (right) models.

\FloatBarrier 
\clearpage

\end{document}